\documentclass[a4paper,12pt]{article}
\usepackage{graphicx}
\usepackage{amssymb}
\usepackage{amsmath}
\usepackage{cite}
\usepackage{axodraw}

\def\be{\begin{equation}}
\def\ee{\end{equation}}
\def\bea{\begin{eqnarray}}
\def\eea{\end{eqnarray}}
\def\beb{\begin{eqnarray*}}
\def\eeb{\end{eqnarray*}}
\def\pat{\partial}
\newcommand{\req}[1]{(\ref{#1})}
\renewcommand{\Re}{{\cal R}}
\newcommand{\newrho}{{\Re}}
\newcommand{\Bmn}{{B}}
\newcommand{\fmn}{{\cal A}}

\newlength{\myVSpace}
\newcommand\xstrut{\raisebox{-.5\myVSpace}
  {\rule{0pt}{\myVSpace}}%
}

\begin{document}
\makeatletter
\def\fmslash{\@ifnextchar[{\fmsl@sh}{\fmsl@sh[0mu]}}
\def\fmsl@sh[#1]#2{%
  \mathchoice
    {\@fmsl@sh\displaystyle{#1}{#2}}%
    {\@fmsl@sh\textstyle{#1}{#2}}%
    {\@fmsl@sh\scriptstyle{#1}{#2}}%
    {\@fmsl@sh\scriptscriptstyle{#1}{#2}}}
\def\@fmsl@sh#1#2#3{\m@th\ooalign{$\hfil#1\mkern#2/\hfil$\crcr$#1#3$}}
\makeatother
\thispagestyle{empty}
\begin{titlepage}


\boldmath
\begin{center}
  {\Large {\bf The Standard Model on Non-Commutative Space-Time:\\[0.3cm]
  Electroweak Currents and the Higgs Sector}}
\end{center}
\unboldmath

\renewcommand{\thefootnote}{\fnsymbol{footnote}}
\begin{center}

{{{\bf
Bla\v zenka Meli\' c${}^1$\footnote{email: melic@thphys.irb.hr},
Kornelija Passek-Kumeri\v{c}ki${}^1$\footnote{email: passek@thphys.irb.hr},\\
Josip Trampeti\'{c}${}^{1,2}$\footnote{email:
josipt@rex.irb.hr},
Peter Schupp${}^3$\footnote{email: p.schupp@iu-bremen.de}\\ and\\
Michael\ Wohlgenannt${}^4$\footnote{email:
michael.wohlgenannt@univie.ac.at} }}}

\end{center}
\setcounter{footnote}{0}
\renewcommand{\thefootnote}{\arabic{footnote}}
\vskip 1em
\begin{center}
${}^{1}$Rudjer Bo\v{s}kovi\'{c} Institute, Theoretical Physics Division, \\
        P.O.Box 180, 10002 Zagreb, Croatia\\
    \vspace{.3cm}
${}^{2}$Max Planck Institut f\"{u}r Physik, \\
F\"{o}hringer Ring 6,
D-80805 M\"{u}nchen, Germany\\
    \vspace{.3cm}
${}^{3}$International University Bremen,\\ Campus Ring 8, 28759 Bremen,
    Germany\\
    \vspace{.3cm}
${}^{4}$Universit\"at Wien, Institut f\"ur theoretische Physik, \\
  Boltzmanngasse 5, 1090 Wien, Austria\\

 \end{center}

\vspace{1cm}

\begin{abstract}
\noindent
In this article we review the electroweak charged and neutral currents
in the Non-Com\-mu\-ta\-tive Standard Mod\-el (NCSM) and
compute the Higgs and Yukawa parts of the NCSM action.
With the aim to make the NCSM accessible
to phenomenological considerations, all relevant expressions are given in
terms of physical fields and Feynman rules are provided.
\end{abstract}

\end{titlepage}

\section{Introduction}
\label{sec:intro}

The approach to non-commutative field theory based on star products
and Seiberg-Witten (SW) maps allows the generalization of the Standard
Model (SM) of particle physics to the case of  non-commutative
space-time, keeping the original gauge group and particle content
\cite{Kontsevich:1997vb,Seiberg:1999vs,Madore:2000en,Jurco:2000ja,
Jurco:2001rq,Calmet:2001na,Chaichian:2004yh,Koch:2004ud}. It
provides a systematic way to compute Lorentz violating operators
that could be a signature of a (hypothetical) non-commutative
space-time structure
\cite{Behr:2002wx,Schupp:2002up,Minkowski:2003jg,Trampetic:2002eb,Schupp:2004dz,
Duplancic:2003hg,Ohl:2004tn,Armoni:2000xr,Bichl:2001cq,
Grimstrup:2002xs,Gomis:2000zz,Brandt:2003fx}.

In this article we carefully discuss the electroweak charged and
neutral currents in the Non-Com\-mu\-ta\-tive Standard Model (NCSM)
\cite{Calmet:2001na} and compute the Higgs and Yukawa parts of the
NCSM action. Among the features which are novel in comparison with
the SM is the appearance of additional gauge boson interaction terms
and of interaction terms without Higgs boson which include
additional mass dependent contributions. All relevant expressions
are given in terms of physical fields and selected Feynman rules are
provided with the aim to make the model more accessible to
phenomenological considerations.

In the star product formulation of non-commutative field theory, one
retains the ordinary functions (and fields) on Minkowski space, but
introduces a new non-commutative product which encodes the
non-commutative structure of space-time. For a constant
antisymmetric matrix $\theta^{\mu\nu}$, the relevant product is the
Moyal-Weyl star product
\begin{equation}
f * g = \sum_{n=0}^\infty \frac{\theta^{\mu_1 \nu_1} \cdots
\theta^{\mu_n \nu_n}}{(-2i)^nn!}
\left( \partial_{\mu_1}\ldots\partial_{\mu_n} f \right)
\left( \partial_{\nu_1}\ldots\partial_{\nu_n} g \right) \,.
\end{equation}
For coordinates: $x^\mu * x^\nu - x^\nu * x^\mu = i
\theta^{\mu\nu}$. More generally, a star product has the form
\begin{equation}
(f * g) (x)= f(x)g(x) + \frac{i}2\theta^{\mu\nu}(x)\pat_\mu f(x) \,
\pat_\nu g(x) + \mathcal O(\theta^2) \, ,
\end{equation}
where the Poisson tensor $\theta^{\mu\nu}(x)$ may be $x$-dependent
and satisfies the Jacobi identity. Higher-order terms in the star
product are chosen in such a way that the overall star product is
associative. In general, they involve derivatives of $\theta$. For a
discussion of the Seiberg-Witten approach to non-commutative field
theory in the case of the space-time dependent $\theta^{\mu\nu}(x)$, see,
for example \cite{Calmet:2003jv,Dimitrijevic:2003wv,Behr:2003qc,
Dimitrijevic:2003pn,Grimstrup:2003rd}.

Carefully studying non-commutative gauge transformations one finds
that in general, non-commutative gauge fields are valued in the
enveloping algebra of the gauge group
\cite{Madore:2000en,Jurco:2000ja}.
(Only for $U(N)$ in the fundamental representation
it is possible to stick to Lie-algebra valued gauge fields.)
A priori this would imply an infinite number
of degrees of freedom if all coefficient functions of the
monomials that form an infinite basis of the enveloping algebra were
independent.
That is the place
where the second important ingredient of gauge
theory on non-commutative spaces comes into play: Seiberg-Witten
maps \cite{Seiberg:1999vs,Madore:2000en} which relate
non-commutative gauge fields and ordinary fields in commutative
theory  via a power series expansion in $\theta$. Since higher-order
terms are now expressed in terms of the zeroth-order fields, we do
have the same number of degrees of freedom as in the commutative
case. Non-commutative fermion and gauge fields read
\begin{eqnarray}
\widehat \psi & = & \widehat \psi[V] = \psi - \, \frac{1}{2}
\, \theta^{\alpha \beta} \,
       V_{\alpha}  \,\partial_{\beta} \psi
          + \frac{i}{8}  \,\theta^{\alpha \beta} \,
              [V_{\alpha}, V_{\beta}]  \,\psi
+ \mathcal O(\theta^2)
\, ,\\
\widehat V_\mu & = & \widehat V_\mu[V] = V_\mu +
    \frac{1}{4} \theta^{\alpha \beta}
           \{ \partial_{\alpha}V_{\mu} + F_{\alpha \mu},V_{\beta}
           \}
+ \mathcal O(\theta^2)
\, ,
\label{eq:hatVmu-exp}
\end{eqnarray}
where $\psi$ and $V_\mu$ are ordinary fermion and gauge fields,
respectively. Non-commutative fields throughout the paper 
are denoted by a hat. The
Seiberg-Witten maps are not unique. The free parameters are chosen
such that the non-commutative gauge fields are hermitian and the
action is real. Still, there is some remaining freedom including the
freedom of the classical field redefinition and the non-commutative gauge
transformation. The noncommutative actions considered here are
covariant under (global) Poincare transformations \emph{provided
that the Poisson tensor $\theta$ is transformed as well.} With
respect to a fixed $\theta$-background, however, the classical Lorentz
symmetry is broken. What remains is a twisted Poincare symmetry
\cite{Chaichian:2004yh,Chaichian:2004za}, which can in principle
be extended to SW expansions.

In \cite{Calmet:2001na}, it was shown how to construct a model with
non-commutative gauge invariance, which stays as close as possible
to the regular Standard Model. The distinguishing feature of this
\emph{minimal} NCSM (mNCSM) is the absence of new triple neutral
gauge boson interactions in the gauge sector. However, as shown
here, triple $Z$ coupling does appear from the Higgs action. Triple
gauge boson interactions do quite naturally arise in the gauge
sector of extended versions
\cite{Calmet:2001na,Behr:2002wx,Duplancic:2003hg, Ohl:2004tn} of the
NCSM and have been discussed in \cite{Behr:2002wx,Duplancic:2003hg}.
They also occur in an alternative approach to the non-commutative
Standard Model given in \cite{Chaichian:2001aa,Chaichian:2004yw}.
Another interesting novel feature of NCSM, introduced by
Seiberg-Witten (SW) maps , is the appearance of mixing of the strong
and electroweak interactions already at the tree level
\cite{Calmet:2001na,Behr:2002wx,Melic:2005am}.

We consider the $\theta$-expanded
NCSM up to first order in the non-co\-mmu\-tat\-ivity parameter
with an emphasis made on the electroweak interactions only.
In Section \ref{sec:NCSM}, we give an introductory overview of the
NCSM. In Section \ref{sec:gauge}, we discuss different choices for
representations of the gauge group which then yield minimal and
non-minimal versions of the NCSM. In Section~\ref{sec:ew}, we carefully
discuss electroweak charged and neutral currents of the NCSM. Explicit
expressions for the NCSM corrections in the Higgs and Yukawa sectors
are worked out in Section~\ref{sec:higgs}. These expressions can be
used directly for further studies. The Feynman rules for the
selected three- and four-field electroweak vertices
are given in Section~\ref{app:ewFR}.


\section{Non-commutative Standard Model}
\label{sec:NCSM}

The action of the NCSM formally resembles the action of the
classical SM: the usual point-wise products in the Lagrangian are
replaced by the Moyal-Weyl product and (matter and gauge) fields are
replaced by the appropriate Seiberg-Witten expansions. In the limit
of vanishing non-commutativity one recovers the usual commutative
theory. This limit is assumed to be continuous. If the transition is
not continuous (compare, e.g. \cite{Armoni:2000xr}), perturbative
aspects of the theory under consideration can still be addressed.
Problems with unitarity may occur in the non-ex\-panded theory with
non-trivial time-space commutation relations. These problems can be
overcome by a careful analysis of perturbation theory in a
Hamiltonian approach, cf. \cite{Gomis:2000zz,Denk:2004um} for scalar
field theory. Other problems in non-commutative theories that are
encountered already at the classical level are charge quantization
in non-commutative QED, the definition of the tensor product of
gauge fields, gauge invariance of the Yukawa couplings and
ambiguities in the kinetic part of the action for gauge fields. As
demonstrated in \cite{Calmet:2001na}, all these problems can be
overcome and do not affect the NCSM presented here. The action of
the NCSM is
\begin{equation}
S_{\mbox{\tiny NCSM}}= S_{\mbox{\tiny fermions}} + S_{\mbox{\tiny gauge}} + S_{\mbox{\tiny{Higgs}}} +
S_{\mbox{\tiny{Yukawa}}}
\, ,
\label{eq:Sncsm}
\end{equation}
where
\begin{eqnarray}
S_{\mbox{\tiny fermions}}& =&
\int d^4x
\sum_{i=1}^3 \left(
\overline{\widehat L}^{(i)}_L
 *   (i\widehat{\fmslash D} \, \widehat L^{(i)}_L)
+  \overline{\widehat Q}^{(i)}_L  *
(i\widehat{\fmslash  D}\,  \widehat Q^{(i)}_L)
\right.
 \nonumber \\ & &
\left.
 + \overline{\widehat e}^{(i)}_R  *
(i\widehat{\fmslash  D}\,  \widehat e^{(i)}_R)
+
 \overline{\widehat u}^{(i)}_R  *
(i\widehat{\fmslash  D}\,  \widehat u^{(i)}_R)
+
 \overline{\widehat d}^{(i)}_R  *
(i\widehat{\fmslash  D}\,  \widehat d^{(i)}_R)
\right)
\, ,
\label{eq:Sfermions}
\end{eqnarray}
\begin{eqnarray}
S_{\mbox{\tiny Higgs}} & = &
 \int d^4x \bigg( h_0^\dagger(\widehat D_\mu \widehat \Phi)
 *  h_0(\widehat D^\mu \widehat \Phi)
- \mu^2 h_0^\dagger(\widehat {\Phi})  *  h_0(\widehat \Phi)
\nonumber \\
&&
- \lambda
h_0^\dagger(\widehat \Phi)  *   h_0(\widehat \Phi)
 *
h_0^\dagger(\widehat \Phi)  *   h_0(\widehat \Phi)   \bigg)
\, ,
\label{eq:Shiggs}
\end{eqnarray}
\begin{eqnarray}
S_{\mbox{\tiny Yukawa}} & = &{\hspace{-2mm}}
 - {\hspace{-2mm}}\int {\hspace{-1mm}}d^4x
\sum_{i,j=1}^3
\nonumber  \\ & &
\bigg (G_e^{(ij)}\,
( \overline{ \widehat L}^{(i)}_L  *  h_{e}(\widehat \Phi)
 *\widehat e^{(j)}_R)
+{G_e^{\dagger}}^{(ij)}\,
( \overline {\widehat e}^{(i)}_R  *  h_{e}(\widehat \Phi)^\dagger
 *\widehat L^{(j)}_L)
\nonumber \\
&&
+
G_u^{(ij)}\,
( \overline{\widehat Q}^{(i)}_L  *  h_{u} (\widehat{\Phi}_c)
 *\widehat u^{(j)}_R)
+{G_u^{\dagger}}^{(ij)} \,
( \overline {\widehat u}^{(i)}_R * h_{u}(\widehat{\Phi}_c)^\dagger
 *\widehat Q^{(j)}_L)
\nonumber \\
&&
+
G_d^{(ij)} \,
( \overline{ \widehat Q}^{(i)}_L  *  h_{d}(\widehat \Phi)
 * \widehat d^{(j)}_R )
+{G_d^{\dagger}}^{(ij)}\,
( \overline{ \widehat d}^{(i)}_R  *  h_{d}(\widehat \Phi)^\dagger
 * \widehat Q^{(j)}_L)
\bigg)
\, .
\nonumber \\ & &
\label{eq:Syukawa}
\end{eqnarray}
The gauge part $S_{\mbox{\tiny gauge}}$ of the action is given in
the next section. The particle spectrum of the SM, as well as
that of the NCSM, is given in Table \ref{tab:SMfields}.
\begin{table}[tp]
\centering
  \begin{tabular}{|c|c|c|c|c|c|}
  \hline \addtolength{\myVSpace}{-.8cm}\xstrut
  & $SU(3)_C$ & $SU(2)_L$ & $U(1)_Y$  & $U(1)_Q$
  & $T_3$
   \\
   \hline
     $ e_R^{(i)}$
   & ${\bf 1}$
   & ${\bf 1}$
   & $-1$
   & $-1$
   & $0$
   \\
  \hline \xstrut
   $ L_L^{(i)}=\left(\begin{array}{c} \nu_L^{(i)} \\ e_L^{(i)} \end{array} \right )$
   & ${\bf 1}$
   & ${\bf 2}$
   & $-1/2$
   & $\left(\begin{array}{c} 0 \\ -1 \end{array} \right )$
   & $\left(\begin{array}{c} 1/2 \\ -1/2 \end{array} \right )$
    \\
  \hline
   $u_R^{(i)}$
   & ${\bf 3}$
   & ${\bf 1}$
   & $2/3$
   & $2/3$
   & $0$
   \\
  \hline
   $d_R^{(i)}$
   & ${\bf 3}$
   & ${\bf 1}$
   & $-1/3$
   & $-1/3$
   & $0$
    \\
  \hline   \xstrut
     $Q_L^{(i)}=\left(\begin{array}{c}  u_L^{(i)} \\ d_L^{(i)} \end{array} \right )$
   & ${\bf 3}$
   & ${\bf 2}$
   & $1/6$
   & $\left(\begin{array}{c} 2/3 \\ -1/3 \end{array} \right )$
   & $\left(\begin{array}{c} 1/2 \\ -1/2 \end{array} \right )$
    \\
  \hline
  \hline\xstrut
$\Phi=\left(\begin{array}{c}  \phi^+ \\  \phi^0 \end{array} \right )$
   & ${\bf 1}$
   & ${\bf 2}$
   & $1/2$
   & $\left(\begin{array}{c} 1 \\ 0 \end{array} \right )$
   & $\left(\begin{array}{c} 1/2 \\ -1/2 \end{array} \right )$
   \\
  \hline
  \hline
$W^+$, $W^-$, $Z$
   & ${\bf 1}$
   & ${\bf 3}$
   & $0$
   & $(\pm 1, 0)$
   & $(\pm 1, 0)$
    \\
  \hline
$A$
   & ${\bf 1}$
   & ${\bf 1}$
   & $0$
   & $0$
   & $0$
    \\
   \hline
$G^b$
   & ${\bf 8}$
   & ${\bf 1}$
   & $0$
   & $0$
   & $0$
   \\
   \hline
 \end{tabular}
\caption{\small The Standard Model fields.
 Here $i\in\{1,2,3\}$ denotes the generation index.
 The electric charge is given by the Gell-Mann-Nishijima relation
 $Q=\left(T_3+Y\right)$.
The physical electroweak fields $A$, $W^+$, $W^-$ and $Z$ are
expressed through the unphysical
$U(1)_Y$ and $SU(2)$ fields
${\cal A}$ and $B_a$ ($a\in\{1,2,3\}$)
in Eq. (\protect\ref{eq:Amu}).
The gluons $G^b$ ($b\in\{1,2,\ldots,8\}$)
are in the octet representation of $SU(3)_C$.
\label{tab:SMfields}}
\end{table}
Analogously to the usual SM definitions for fermion
fields, we define $\overline{\widehat{\psi}}=\widehat{\psi}^{\dagger}\,\gamma^0$.
(The $\gamma$ matrix can be pulled out of the SW expansion because
it commutes with the matrices representing internal symmetries.)
The indices $L$ and $R$ denote the standard left and right components
$\psi_L=1/2 (1-\gamma_5)\psi$ and $\psi_R=1/2 (1+\gamma_5)\psi$. For
the conjugate Higgs field, we have $\Phi_c = i \tau_2 \Phi^*$
($\tau_2$ is the usual Pauli matrix). In Eqs.~\req{eq:Sfermions} and
\req{eq:Syukawa} the generation index is denoted by $i,j \in
\{1,2,3\}$. The matrices $G_e$, $G_u$ and $G_d$ are the Yukawa couplings.

The non-commutative Higgs field $\widehat{\Phi}$ is given by the
hybrid SW map
\begin{eqnarray}
\widehat{\Phi} &\equiv& \widehat{\Phi}[\Phi,V,V']
   \nonumber \\
  &=& \Phi +
  \frac{1}{2} \, \theta^{\alpha \beta}
     V_{\beta}
  \left( \partial_{\alpha} \Phi
       - \frac{i}{2} (V_{\alpha} \Phi - \Phi V'_{\alpha}) \right)
     \\ & &
   + \frac{1}{2} \, \theta^{\alpha \beta}
  \left( \partial_{\alpha} \Phi
       - \frac{i}{2} (V_{\alpha} \Phi - \Phi V'_{\alpha}) \right)
       V'_{\beta}
       + \mathcal O(\theta^2)
     \nonumber
\, ,
\end{eqnarray}
which generalizes the Seiberg-Witten maps of both gauge bosons and fermions.
$\widehat\Phi$ is a functional of two gauge fields $V$ and $V'$ and
transforms covariantly under gauge transformations:
\begin{equation}
\label{phi-delta}
\delta \widehat\Phi[\Phi,V,V'] = i\widehat \Lambda * \widehat \Phi - i\widehat
\Phi*\widehat \Lambda'\, ,
\end{equation}
where $\widehat\Lambda$ and $\widehat\Lambda'$ are the corresponding gauge
parameters. Hermitian conjugation yields
$\widehat\Phi[\Phi,V,V']^\dagger = \widehat\Phi[\Phi^\dagger,V',V]$.
The covariant derivative for the non-commutative Higgs field
$\widehat{\Phi}$ is given by
\begin{equation}
\widehat{D}_{\mu} \widehat{\Phi} =
\partial_{\mu} \widehat{\Phi}
-i \, \widehat{V}_{\mu} * \widehat{\Phi}
+i \, \widehat{\Phi} * \widehat{V}'_{\mu}
\, .
\label{eq:hatDPhi}
\end{equation}
As explained in \cite{Calmet:2001na},
the precise representations of the gauge fields
$V$ and $V'$ in the Yukawa couplings
are inherited from the
fermions
on the left ($\bar{\psi}$)
and on the right side ($\psi$)
of the Higgs field found in \req{eq:Syukawa}, respectively.
The following notation was introduced
in Eqs.~\req{eq:Shiggs} and \req{eq:Syukawa}
\begin{eqnarray}
h_0(\widehat{\Phi} ) &=& \widehat{\Phi}[\Phi,\frac{1}{2} g' {\cal A}
+ g B^a T_{L}^a,0]
\, ,
\nonumber \\
h_{\psi}(\widehat{\Phi} ) &=& \widehat{\Phi}[\Phi, {\newrho}_{\psi_L}(V),
    {\newrho}_{\psi_R}(V)]
\, ,
\label{eq:def-h}\\
h_{\psi}(\widehat{\Phi}_c ) &=& \widehat{\Phi}[\Phi_c, {\newrho}_{\psi_L}(V),
{\newrho}_{\psi_R}(V)]
\, .\nonumber
\end{eqnarray}
The representations ${\newrho}_{\psi}$,
determined by the multiplet $\psi$,
are listed in Table \ref{tab:rhoPsi}.
\begin{table}[htp]
\centering
  \begin{tabular}{|c|c|}
   \hline\addtolength{\myVSpace}{-.6cm}\xstrut
   $\psi$ & ${\newrho}_{\psi}(V_\mu)$
   \\
    \hline\hline\addtolength{\myVSpace}{-.6cm}\xstrut
   $ e_R^{(i)}$
   & $-g'\, {\cal A}_\mu$   \\
  \hline\xstrut
     $ L_L^{(i)}=\left(\begin{array}{c} \nu_L^{(i)} \\ e_L^{(i)} \end{array} \right )$
   & $-\frac{1}{2} \, g'\, {\cal A}_\mu+g\, B_{\mu}^a \, T^a_L$
   \\
  \hline\addtolength{\myVSpace}{-.6cm}\xstrut
   $u_R^{(i)}$
   & $\frac{2}{3} \, g'\, {\cal A}_\mu
  +g_s \, G_{\mu}^{b} \, T^b_S$ \\
  \hline\addtolength{\myVSpace}{-.6cm}\xstrut
   $d_R^{(i)}$
   & $-\frac{1}{3}\,  g'\, {\cal A}_\mu
  +g_s \, G_{\mu}^{b} \, T^b_S$ \\
   \hline \xstrut
   $Q_L^{(i)}=\left(\begin{array}{c}  u_L^{(i)} \\ d_L^{(i)} \end{array} \right )$
   & $\frac{1}{6}\,  g' \, {\cal A}_\mu+g\,   B_{\mu}^{a} \, T^a_L
  +g_s\,  G_{\mu}^{b} \, T^b_S$   \\
  \hline
 \end{tabular}
\caption{\small
The gauge fields
in the covariant derivatives of the fermions and
in the Seiberg-Witten maps of the fermions
in the Non-Commutative Standard Model.
The matrices $T^a_L= \tau^a/2$ and
$T^b_S=\lambda^b/2$ correspond to the Pauli and Gell-Mann matrices
respectively, and the summation over the indices
 $a\in\{1,2,3\}$ and $b\in\{1,\ldots,8\}$ is understood.
\label{tab:rhoPsi}}
\end{table}
Note that
${\newrho}_{\psi}(f(V_{\mu}))=f({\newrho}_{\psi}(V_{\mu}))$
for any function $f$.
Gauge invariance does not restrict the choice of representation for
the Higgs field in $S_{\mbox{\tiny Higgs}}$. The simplest choice for $h_0$
which is adopted in the NCSM closely follows the SM representation
for the Higgs field.
For a better understanding of the gauge
invariance, let us consider the hypercharges in two examples: 
\begin{equation}
\begin{array}{rccccc}
 & \overline{\widehat L}_L[V] & * &
 \widehat\Phi[\Phi,V,V'] & * &
    \widehat e_R[V']
    \\[1ex]
{ Y : \quad} & 1/2& &\underbrace{{ -1/2 }\, ;\, { 1}}_{1/2} && -1\,,
\\[4ex]
&\overline{\widehat Q}_L[V] & * & \widehat\Phi[\Phi,V,V'] & * &
    \widehat d_R[V']\\
[1ex]
{ Y : \quad} & -1/6 &  &\underbrace{{ 1/6}\, ;\, { 1/3}}_{1/2} && -1/3\,.
\end{array}
\end{equation}
The choice of
representation allows us to assign
separate left and right hypercharges to the noncommutative Higgs field
$\widehat\Phi$, which add up to Higgs usual hypercharge \cite{Calmet:2001na}.
Because of the minus sign in (\ref{phi-delta}), the right
hypercharge attributed to the Higgs is effectively $-Y_{\psi_R}$.

In Grand Unified Theories (GUT) it is more natural to first combine
the left-handed and right-handed
fermion fields and then contract the resulting expression
with Higgs fields to obtain a gauge-invariant Yukawa term.
Consequently, in NC GUTs we need to use the hybrid SW map for the left-handed
fermion fields and then sandwich them between the NC Higgs on the left- and
the right-handed fermion fields on the right  \cite{Aschieri:2002mc}.

\section{\label{sec:gauge}Gauge Sector of the NCSM Action}

The general form of the gauge kinetic terms is
\cite{Aschieri:2002mc}
\bea
S_{\mathrm{gauge}} = -\frac{1}{2}  \int
d^4x \, \sum_{\newrho} {c_{\newrho}} {\rm \bf Tr}\Big(
{\newrho}(\widehat F_{\mu \nu}) * {\newrho}(\widehat F^{\mu
\nu})\Big), \label{action1}
\eea
where the non-commutative field strength $\widehat{F}_{\mu \nu}$
\begin{eqnarray}
\widehat{F}_{\mu \nu}&=&\partial_{\mu} \widehat{V}_{\nu}
            -\partial_{\nu} \widehat{V}_{\mu}
            -i [\widehat{V}_{\mu} \stackrel{*}{,} \widehat{V}_{\nu}]
           \nonumber \\
    &=& F_{\mu \nu} +
    \frac{1}{2} \theta^{\alpha \beta}
           \{ F_{\mu \alpha}, F_{\nu \beta} \}
                 - \frac{1}{4} \theta^{\alpha \beta}
           \{ V_{\alpha}, (\partial_{\beta}+D_{\beta}) F_{\mu \nu} \}
\, + {\cal O}(\theta^2)
\, ,
\label{eq:hatFmunu-exp}
\end{eqnarray}
was obtained  from
the SW map for the non-commutative
vector potential (\ref{eq:hatVmu-exp}).
Ordinary field strength $F_{\mu \nu}$ is given by
\begin{equation}
F_{\mu \nu}=\partial_{\mu} V_{\nu}-\partial_{\nu} V_{\mu}
            -i [V_{\mu},V_{\nu}]
\, ,
\label{eq:Fmunu}
\end{equation}
while its covariant derivative reads
\begin{equation}
D_{\beta} F_{\mu \nu}=\partial_{\beta} F_{\mu \nu}
                -i [V_{\beta},F_{\mu \nu}]
\, .
\label{eq:DFmunu}
\end{equation}
Here $V_{\mu}$ represents
the whole of the gauge potential for the SM gauge group,
\begin{equation}
V_{\mu}(x)=g' {\cal A}_{\mu}(x) Y
          + g \sum_{a=1}^{3} B_{\mu}^a(x) \, T_{L}^a
          + g_s \sum_{b=1}^{8} G_{\mu}^{b}(x) \, T_S^b
\, .
\label{eq:Vmu}
\end{equation}

The sum in (\ref{action1}) is over all unitary,
irreducible and inequivalent representations ${\newrho}$ of a
gauge group. The freedom in the kinetic terms is parametrized
by real coefficients $c_{\newrho}$ that are subject to the
constraints
\bea
\frac{1}{g^2_{I}} = \sum_{\newrho} c_{\newrho}
{\rm \bf Tr}\Big( {\newrho}(T^a_I) {\newrho}(T^a_I)\Big),
\label{constr-action2}
\eea
where $g_{I}$ are the usual
``commutative'' coupling constants $g'$, $g$, $g_s$
and $T^a_I$ are generators of $U(1)_Y$,
$SU(2)_L$, $SU(3)_C$, respectively.
Equations (\ref{action1}) and (\ref{constr-action2}) 
can also be written more compactly as
\begin{equation}
\label{action1a}
S_{\mathrm{gauge}} = -\frac{1}{2}  \int d^4x \,
\mbox{\bf Tr} \frac{1}{\mbox{\bf G}^2} \widehat F_{\mu \nu} * \widehat F^{\mu \nu},
\qquad
\frac{1}{g^2_{I}} = \mbox{\bf Tr} \frac{1}{\mbox{\bf G}^2} T^a_I T^a_I,
\end{equation}
where the trace \mbox{\bf Tr} is again over all representations
and \mbox{\bf G} is an operator that commutes with all
generators $T^a_I$ and encodes the coupling constants \cite{Behr:2002wx}.
The trace  in the kinetic terms for gauge bosons is not unique,
it depends on the choice of representation.
This would not be of importance if the gauge fields
were Lie algebra valued, but in the noncommutative case
they live in the enveloping algebra.
The possibility of new parameters in gauge theories on non-commutative
space-time is a consequence of the fact that the gauge fields can
take any value in the enveloping algebra of the gauge group.

It is
instructive to provide the general form of $S_{\mbox{\tiny gauge}}$,
(\ref{action1}), in terms of SM fields:
\begin{eqnarray}
S_{\mbox{\tiny gauge}} &=&
- \frac{1}{2} \int d^4 x \,
\mbox{\bf Tr} \frac{1}{\mbox{\bf G}^2} F_{\mu \nu} F^{\mu \nu}
\nonumber \\ & &
+
\theta^{\rho \sigma} \int d^4 x \,
\mbox{\bf Tr} \frac{1}{\mbox{\bf G}^2}
\left[ \left( \frac{1}{4} F_{\rho \sigma} F_{\mu \nu}
-F_{\rho \mu} F_{\sigma \nu} \right) F^{\mu \nu} \right]
+ {\cal O}(\theta^2)
\, . \quad
\label{eq:genSgaugeSM}
\end{eqnarray}

\subsection{Minimal NCSM}

In the minimal Non-Commutative Standard Model (mNCSM)
which adopts
the whole of the gauge potential (\ref{eq:Vmu}) for the SM gauge group,
the mNCSM gauge action is given by
\bea
S^{\mbox{\tiny mNCSM}}_{\mbox{\tiny gauge}}& =& -\frac{1}{2} \int
d^4x \left( \frac{1}{{g'}^2} \mbox{Tr}_{\bf 1} + \frac{1}{g^2}
\mbox{Tr}_{\bf 2} +\frac{1}{g_s^2} \mbox{Tr}_{\bf 3}
\right) \widehat F_{\mu \nu}  *   \widehat F^{\mu \nu} \, .
\label{eq:SmNCSM} \eea
Here the simplest choice was taken, i.e, a sum of three
traces over the $U(1)$, $SU(2)$, $SU(3)$ sectors
with
 \begin{equation}\label{2}
Y = \frac{1}{2} \left(\begin{array}{rr} 1 & 0 \\ 0 & -1\end{array}\right)
\, ,
\end{equation}
in the definition of $\mbox{Tr}_{\bf 1}$
and the
fundamental representations for $SU(2)$ and $SU(3)$ generators
in $\mbox{Tr}_{\bf 2}$
and $\mbox{Tr}_{\bf 3}$, respectively.
In terms of physical fields, the action then reads
\bea
S^{\mbox{\tiny mNCSM}}_{\mbox{\tiny gauge}} &=&
-\frac12 \int d^4x \left( \frac12 \fmn_{\mu\nu}\fmn^{\mu\nu}
+ \, \mbox{Tr}\, {\Bmn}_{\mu\nu} {\Bmn}^{\mu\nu}
+ \, \mbox{Tr}\, G_{\mu\nu} G^{\mu\nu} \right)
\label{eq:SmNCSMa}\\
& & + \frac14 \, g_s\, {d^{abc}} \, \theta^{\rho\sigma}
\int d^4x
\left(
\frac{1}{4}G^a_{\rho\sigma} G^b_{\mu\nu}
-G^a_{\rho\mu} G^b_{\sigma\nu}
\right)G^{\mu\nu,c} + \, {\cal O}(\theta^2) \, ,
\nonumber
\eea
where $\fmn_{\mu\nu}$, ${\Bmn}_{\mu\nu}(={\Bmn}_{\mu\nu}^aT_L^a)$
and $G_{\mu\nu}(=G_{\mu\nu}^aT_S^a)$ denote the $U(1)$, $SU(2)_L$ and $SU(3)_c$ field strengths,
respectively:
\bea
\fmn_{\mu\nu}&=&\partial_{\mu}{\cal A}_{\nu}-\partial_{\nu}{\cal A}_{\mu}\:,
\nonumber\\
{\Bmn}_{\mu\nu}^a &=& \partial_{\mu}B^a_{\nu}-\partial_{\nu}B^a_{\mu}
+g\;\epsilon^{abc}B^b_{\mu} B^c_{\nu}\:,
\nonumber\\
G_{\mu\nu}^a &=& \partial_{\mu}G^a_{\nu}-\partial_{\nu}G^a_{\mu}
+g_s\;f^{abc}G^b_{\mu} G^c_{\nu}\:.
\label{eq:fFG}
\eea
Note that in order
to obtain the above result%
\footnote{Note that hereby we correct Eq. (56)
of Ref. \cite{Calmet:2001na}.},
one makes use of  the following symmetry properties of the group generators
$T_{L}^a=\tau^a/2$ and $T_S^a=\lambda^a/2$:
$$
\mbox{Tr}(T^a T^b) = \frac12 \delta^{ab}
\, , \quad
\mbox{Tr} (\tau^a \tau^b \tau^c)=2i\epsilon^{abc}
\, , \quad
\mbox{Tr} (\lambda^a \lambda^b \lambda^c)=2(d^{abc} + if^{abc})
\, ,
$$
where $\epsilon^{abc}$ is the usual antisymmetric tensor,
while $f^{abc}$ and $d^{abc}$ are totally antisymmetric
and totally symmetric structure constants of the $SU(3)$ group.

There are no new electroweak gauge boson
interactions in Eq.~(\ref{eq:SmNCSMa})
nor
the vertices already present
in SM, like $W^+W^-\gamma$ and $W^+W^-Z$,
do acquire any corrections.
This is a consequence of
our choice of the hypercharge (\ref{2}) and
of the antisymmetry in both the Lorentz and the group representation
indices.
However, new couplings, like $ZZZ$,
and $\theta$ corrections to SM vertices
enter from the Higgs kinetic terms as elaborated in Section \ref{H}.

For the convenience of the reader,
we list some usual definitions
that we use in the analysis
of the electroweak sector.
The physical fields for the electroweak gauge bosons ($W^{\pm}$, $Z$)
and the photon ($A$) are given by
\begin{eqnarray}
W^{\pm}_{\mu}&=&
\frac{B_{\mu}^1\mp i B_{\mu}^2}{\sqrt{2}}
\, ,
\nonumber\\
Z_{\mu} &=&
\frac{-g' {\cal A}_{\mu} + g B_{\mu}^3}{\sqrt{g^2+{g'}^2}}
        = -\sin \theta_W {\cal A}_{\mu}
            + \cos \theta_W B_{\mu}^3
        \, ,
\nonumber\\
A_{\mu} &=& \frac{g {\cal A}_{\mu} + g' B_{\mu}^3}{\sqrt{g^2+{g'}^2}}
        = \cos \theta_W {\cal A}_{\mu}
            + \sin \theta_W B_{\mu}^3
\, ,
\label{eq:Amu}
\end{eqnarray}
where electric charge $e=g \sin \theta_W=g' \cos \theta_W$.

\subsection{\label{sec:non-minimal}Non-Minimal NCSM}

We can use the freedom in the choice of
traces in kinetic terms for gauge fields
to construct non-minimal versions of the mNCSM (nmNCSM).
Since the fermion-gauge boson interactions remain the same
regardless on the choice of traces in the gauge sector,
the matter sector of the action is not affected,
i.e. it is the same for both versions of the NCSM.

The expansion in $\theta$ is at the same time an expansion in the
momenta. The $\theta$-expanded action can thus be interpreted as a
low-energy effective action. In such an effective low-energy
description it is natural to expect that all representations that
appear in commutative theory (matter multiplets and adjoint
representation) are important.
All representations
of gauge fields that appear in the SM then have to be considered in
the definition of the trace \req{action1a}.
In \cite{Behr:2002wx}
the trace was chosen over all particles
on which covariant derivatives act
and which have different quantum numbers.
In the SM, these are, five multiplets
of fermions for each generation and one Higgs multiplet.
The operator ${\bf G}$, which determines the coupling constants of the theory,
must commute with all generators
$(Y,T^a_L,T^b_S)$ of the gauge group,
so that it does not spoil the trace property of ${\bf Tr}$.
This implies that ${\bf G}$
takes on constant values $g_1,\ldots,g_6$
on the six multiplets (Table \ref{tab:SMfields}).
The operator ${\bf G}$ is in general a function of $Y$
and of the Casimir  operators of $SU(2)$ and $SU(3)$.
The action derived from (\ref{eq:genSgaugeSM})
for such nmNCSM takes the following form:
\begin{eqnarray}
S_{\mbox{\tiny gauge}}^{\mbox{\tiny nmNCSM}}
&=& S^{\mbox{\tiny mNCSM}}_{\mbox{\tiny gauge}}
\nonumber \\
&&+{g'}^3\kappa_1{\theta^{\rho\sigma}}\hspace{-2mm}\int \hspace{-1mm}d^4x\,
\left(\frac{1}{4}\fmn_{\rho\sigma}\fmn_{\mu\nu}-\fmn_{\mu\rho}
\fmn_{\nu\sigma}\right)\fmn^{\mu\nu}
\nonumber \\
&&+g'g^2\kappa_2 \, \theta^{\rho\sigma}\hspace{-2mm}\int
\hspace{-1mm} d^4x
\left[(\frac{1}{4}\fmn_{\rho\sigma} {\Bmn}^a_{\mu\nu}-
\fmn_{\mu\rho}{\Bmn}^a_{\nu\sigma}){\Bmn}^{\mu\nu,a}\!+c.p.\right]
 \nonumber \\
&&+g'g^2_s\kappa_3\, \theta^{\rho\sigma}\hspace{-2mm}\int
\hspace{-1mm} d^4x
\left[(\frac{1}{4}\fmn_{\rho\sigma}G^b_{\mu\nu}-
\fmn_{\mu\rho}G^b_{\nu\sigma})G^{\mu\nu,b}\!+c.p.\right]
\nonumber \\
&&
+{\cal O}(\theta^2) \, ,
\label{eq:nmNCSM}
\end{eqnarray}
where $c.p.$ denotes cyclic permutations
of field strength tensors with respect to Lorentz indices.
The constants $\kappa_1$, $\kappa_2$ and $\kappa_3$
represent parameters
of the model given in \cite{Behr:2002wx,Duplancic:2003hg}.
In the following we comment only the pure triple
electroweak gauge-boson interactions.

New anomalous triple-gauge boson interactions that are usually
forbidden by Lorentz invariance, angular moment conservation and
Bose statistics (Landau-Pomeranchuk-Yang theorem) can arise within
the framework of the nmNCSM \cite{Behr:2002wx,Duplancic:2003hg}, but
also in the alternative approach to the NCSM given in
\cite{Chaichian:2001aa}. Neutral
triple-gauge boson terms which are not present in the SM Lagrangian
can be extracted from the action (\ref{eq:nmNCSM}).
In terms of physical fields
($A,Z$) they are
\begin{eqnarray}
{\cal L}_{\gamma\gamma\gamma}&=&\frac{e}{4} \sin2{\theta_W}\;{\rm K}_{\gamma\gamma\gamma}
{\theta^{\rho\sigma}}A^{\mu\nu}\left(A_{\mu\nu}A_{\rho\sigma}-4A_{\mu\rho}A_{\nu\sigma}\right)
\, ,
\nonumber
\\[0.2cm]
{\cal L}_{Z\gamma\gamma}&=&\frac{e}{4} \sin2{\theta_W}\,{\rm K}_{Z\gamma \gamma}\,
{\theta^{\rho\sigma}}
\left[2Z^{\mu\nu}\left(2A_{\mu\rho}A_{\nu\sigma}-A_{\mu\nu}A_{\rho\sigma}\right)\right.\nonumber\\
& & +\left. 8 Z_{\mu\rho}A^{\mu\nu}A_{\nu\sigma} - Z_{\rho\sigma}A_{\mu\nu}A^{\mu\nu}\right]
\, ,
\nonumber
\\[0.2cm]
{\cal L}_{ZZ\gamma}&=&{\cal L}_{Z\gamma\gamma}(A_{\mu}\leftrightarrow Z_{\mu}
)
\, ,
\nonumber
\\
{\cal L}_{ZZZ}&=&{\cal L}_{\gamma\gamma\gamma}(A_{\mu}\to Z_{\mu}
)
\, ,
\label{L3456}
\end{eqnarray}
where
\begin{eqnarray}
{\rm K}_{\gamma\gamma\gamma}&=&\frac{1}{2}\; gg'(\kappa_1 + 3 \kappa_2)
\, ,  \nonumber \\
{\rm K}_{Z\gamma\gamma}&=&\frac{1}{2}\; \left[{g'}^2\kappa_1 + \left({g'}^2-2g^2\right)\kappa_2\right]\,,
\nonumber\\
{\rm K}_{ZZ\gamma}&=&\frac{-1}{2gg'}\; \left[{g'}^4\kappa_1 + g^2\left(g^2-2{g'}^2\right)\kappa_2\right]\,,
\nonumber\\
{\rm K}_{ZZZ}&=&\frac{-1}{2g^2}\; \left({g'}^4\kappa_1 + 3g^4\kappa_2\right)\,,
\label{K123456}
\end{eqnarray}
and here we have introduced the shorthand notation
$X_{\mu\nu}\equiv\partial_{\mu}X_{\nu}-\partial_{\nu}X_{\mu}$
for $X \in \{A,Z\}$.
Details of the derivations of neutral triple-gauge boson terms and
the properties of the coupling constants
in (\ref{eq:nmNCSM})
are explained in \cite{Behr:2002wx,Duplancic:2003hg}.

Additionally, in contrast to the mNCSM \req{eq:SmNCSMa}, electroweak
triple-gauge boson terms already present in the SM acquire  $\theta$
corrections in the nmNCSM. Such contributions which originate from
\req{eq:nmNCSM} read
\begin{eqnarray}
{\cal L}_{WW\gamma}&=&
{\cal L}_{WW\gamma}^{\mbox{\tiny{SM}}}+
{\cal L}_{WW\gamma}^{\theta}
+ {\cal O}(\theta^2) \, ,
\nonumber \\
{\cal L}_{WWZ}&=&
{\cal L}_{WWZ}^{\mbox{\tiny{SM}}}+
{\cal L}_{WWZ}^{\theta}
+ {\cal O}(\theta^2) \, ,
\nonumber \\
{\cal L}_{WW\gamma}^{\theta}&=&\frac{e}{2} \sin2{\theta_W}\,{\rm K}_{WW\gamma}\,
{\theta^{\rho\sigma}}
\left\{A^{\mu\nu}\left[2\left(W^+_{\mu\rho}W^-_{\nu\sigma}
+W^-_{\mu\rho}W^+_{\nu\sigma}\right)
\right. \right.
\nonumber \\
& &
\left. \left.
- \left(W^+_{\mu\nu}W^-_{\rho\sigma}+W^-_{\mu\nu}W^+_{\rho\sigma}\right)
\right]
+ 4 A_{\mu\rho}\left[W^{+\mu\nu}W^-_{\nu\sigma}
+W^{-\mu\nu}W^+_{\nu\sigma}\right]
\right.\nonumber\\
& &
\left.
- A_{\rho\sigma}W^+_{\mu\nu}W^{-\mu\nu}\right\}
\, ,
\nonumber \\[0.2cm]
{\cal L}_{WWZ}^{\theta}&=&
{\cal L}_{WW\gamma}^{\theta}(A_{\mu}\rightarrow Z_{\mu}
)
\, ,
\label{LWW}
\end{eqnarray}
with
\begin{eqnarray}
{\rm K}_{WW\gamma}&=&
-\frac{g}{2 g'}\left[{g'}^2+g^2\right]\kappa_2 \, ,
\nonumber \\[0.2cm]
{\rm K}_{WWZ}&=&-\frac{g'}{g}{\rm K}_{WW\gamma}
\, .
\label{WWgammaZ}
\end{eqnarray}
It is important to stress that in both the mNCSM and the nmNCSM
there are additional $\theta$ corrections to these vertices coming
from the Higgs part of the action. This will be elaborated in detail
in Section \ref{H}.

The new parameters
in the non-minimal NCSM can be restricted by considering GUTs on
non-commutative space-time \cite{Aschieri:2002mc}.


\section{Electroweak Matter Currents}
\label{sec:ew}
In this section we concentrate on the fermion electroweak sector of
the NCSM. Some terms are derivative valued. Nevertheless, the
hermiticity of the Seiberg-Witten maps for the gauge field
guarantees the reality of the action.
Using the SW maps
of the non-commutative fermion field $\widehat{\psi}$ with corresponding function
${\newrho}_{\psi}(V_{\alpha})$
\begin{eqnarray}
\widehat{\psi}=\psi - \, \frac{1}{2} \, \theta^{\alpha \beta} \,
       {\newrho}_{\psi}(V_{\alpha})  \,\partial_{\beta} \psi
          + \frac{i}{8}  \,\theta^{\alpha \beta} \,
              [{\newrho}_{\psi}(V_{\alpha}),{\newrho}_{\psi}(V_{\beta})]  \,\psi
           + {\cal O}(\theta^2)
\, ,
\label{eq:Psi1}
\end{eqnarray}
and it's covariant derivative
\begin{eqnarray}
\widehat{D}_{\mu} \widehat{\psi} &=&
             \partial_{\mu} \widehat{\psi}
             - i {\newrho}_{\psi}(\widehat{V}_{\mu}) * \widehat{\psi}
\nonumber \\
&=& D_{\mu} \left[\psi - \frac{1}{2} \theta^{\alpha \beta} \,
       {\newrho}_{\psi}(V_{\alpha})  \,\partial_{\beta} \psi
          + \frac{i}{8}  \,\theta^{\alpha \beta} \,
              [{\newrho}_{\psi}(V_{\alpha}),{\newrho}_{\psi}(V_{\beta})]\psi \right]
\label{eq:hatDPsi-exp}\\
&-& i \, {\newrho}_{\psi}
\left(
\frac{1}{4} \theta^{\alpha \beta}
           \{ \partial_{\alpha}V_{\mu} + F_{\alpha \mu},V_{\beta}
           \}
\right) \, \psi
 + \frac{1}{2} \, \theta^{\alpha \beta} \,
   (\partial_{\alpha} {\newrho}_{\psi}(V_{\mu})) \, \partial_{\beta} \psi
           + {\cal O}(\theta^2)
\, ,
\nonumber
\end{eqnarray}
it is straightforward to derive  the general expression
\begin{eqnarray}
S_{\mbox{\tiny $\psi$}} & = & \int d^4x \, \overline{\widehat \psi} \,
 *  i \widehat{\fmslash D} \, \widehat \psi
 \nonumber\\
& = & \int d^4x
\left( i \overline{\psi} \, \fmslash D \, \psi
-\frac{i}{4}\,  \overline{\psi}\,
 \theta^{\mu \nu \rho} \, {\newrho}_{\psi}(F_{\mu \nu}) \, D_{\rho} \psi
+ {\cal O}(\theta^2) \right)
\, ,
\label{eq:hatPsiDPsi}
\end{eqnarray}
where $\theta^{\mu \nu \rho}$ is a totally antisymmetric quantity:
\begin{eqnarray}
\theta^{\mu \nu \rho}=
\theta^{\mu \nu} \gamma^{\rho}
+ \theta^{\nu \rho} \gamma^{\mu}
+ \theta^{\rho \mu} \gamma^{\nu}\,.
\label{eq:theta3}
\end{eqnarray}
The terms of the form given in
Eq. \req{eq:hatPsiDPsi} appear in $S_{\mbox{\tiny fermions}}$
\req{eq:Sfermions}. One can easily show that $S^{\dagger}_{\mbox{\tiny
fermions}}=S_{\mbox{\tiny fermions}}$, to order
${\cal O}(\theta^2)$. From Eq.~\req{eq:hatPsiDPsi}
we have
\begin{displaymath}
S_{\mbox{\tiny $\psi$}}^{\dagger}=S_{\mbox{\tiny $\psi$}}-
\frac{i}{4} \,\int \, d^4x \,
\left(
 \overline{\psi}\,
 \theta^{\mu \nu \rho} \;
{\newrho}_{\mbox{\tiny $\psi$}} ( D_{\rho} F_{\mu \nu} ) \,  \psi
\right)
 + {\cal O}({\theta^2})
\, .
\end{displaymath}
Since
${\newrho}_{\mbox{\tiny $\psi$}}(\theta^{\mu \nu \rho} D_{\rho} F_{\mu \nu})=
\theta^{\mu \nu \rho}{\newrho}_{\mbox{\tiny $\psi$}}(D_{\rho} F_{\mu \nu})$
for constant $\theta$, and
\begin{displaymath}
 \theta^{\mu \nu \rho} \, ( D_{\rho} F_{\mu \nu})
= \theta^{\mu \nu} \, \gamma^{\rho}
( D_{\rho} F_{\mu \nu} +
 D_{\nu} F_{\rho \mu} +
 D_{\mu} F_{\nu \rho}  )
\, ,
\end{displaymath}
the $\theta$-dependent term vanishes due to the Bianchi identity
\begin{displaymath}
D_\rho F_{\mu\nu} + D_\nu F_{\rho\mu} + D_\mu F_{\nu\rho} = 0
\, ,
\end{displaymath}
thereby proving the reality of the action $S_{\mbox{\tiny $\psi$}}$
and, hence, the reality of the action $S_{\mbox{\tiny fermions}}$ to
${\cal O}(\theta^2)$. However, note that the reality of the action
is not essential, but is very desirable. \footnote{Weinberg writes
in his book: ``The action is supposed to be real. This is because we
want just as many field equations as there are fields. [\ldots] The
reality also ensures that the generators of various symmetry
transformations are Hermitian operators." \cite{Weinberg:1995mt}}

Next, we express the NCSM results
for the electroweak currents
in terms of physical fields
starting with the left-handed electroweak
sector.
In the following $\Psi_L$ represents $\Psi_L \in \{L_L^{(i)},Q_L^{(i)}\}$
and has the general form
\begin{equation}
\Psi_L = \left(
\begin{array}{c}
\psi_{\mbox{\tiny up,L}} \\
\psi_{\mbox{\tiny down,L}}
\end{array}
\right)
\, .
\end{equation}
In this case, according to the Table \ref{tab:rhoPsi}
the representation ${\newrho}_{\Psi_L}(V_{\mu})$
without $SU(3)$ fields
takes the form
\begin{equation}
{\newrho}_{\Psi_L}(V_{\mu})= g' \, {\cal A}_{\mu} \, Y_{\Psi_L} +
g\, B_{\mu}^a T_L^{a} \, .\label{eq:left}
\end{equation}
The hypercharge generator $Y_{\Psi_L}$ (see Table \ref{tab:SMfields}) can be rewritten as
\begin{equation}
Y_{\Psi_L}=Q_{\psi_{\mbox{\tiny up}}}-T_{3,\psi_{\mbox{\tiny up},L}}
        =Q_{\psi_{\mbox{\tiny down}}}-T_{3,\psi_{\mbox{\tiny down},L}}
\, ,
\end{equation}
and we make use of Eqs. (\ref{eq:Amu}).
The left-handed electroweak part
of the action $S_{\mbox{\tiny $\psi$}}$
can be cast in the form
\begin{eqnarray}
S_{\mbox{\tiny $\psi$,ew,L}}
 &=&
 \int d^4 x \left( \bar\Psi_L\, i\fmslash \partial\, \Psi_L +
\bar\Psi_L \, {\mathbf J}^{(L)} \, \Psi_L \right)
\nonumber \\
 &=&
 \int d^4 x \left( \bar\Psi_L\, i\fmslash \partial\, \Psi_L
+
\bar\psi_{\mbox{\tiny up},L} \, J_{12}^{(L)} \, \psi_{\mbox{\tiny down},L}
+
\bar\psi_{\mbox{\tiny down},L} \, J_{21}^{(L)} \, \psi_{\mbox{\tiny up},L}
\right.
\nonumber \\
 &&
\left.
\, + \,
\bar\psi_{\mbox{\tiny up},L} \, J_{11}^{(L)} \, \psi_{\mbox{\tiny up},L}
+
\bar\psi_{\mbox{\tiny down},L} \, J_{22}^{(L)} \, \psi_{\mbox{\tiny down},L}
\right)
\, ,
\label{eq:SewL}
\end{eqnarray}
where ${\mathbf J^{(L)}}$
is a $2\times 2$ matrix whose
off-diagonal elements ($J_{12}^{(L)}$, $J_{21}^{(L)}$)
denote
the charged currents
and diagonal elements ($J_{11}^{(L)}$, $J_{22}^{(L)}$)
the neutral currents.
After some algebra we obtain
\begin{subequations}
\begin{eqnarray}
J_{12}^{(L)} &=& \frac{g}{\sqrt{2}} \fmslash W^+
           + J_{12}^{(L,\theta)}
           + {\cal O}(\theta^2)
\, ,
\label{eq:JL12}
\\[0.2cm]
J_{21}^{(L)} &=& \frac{g}{\sqrt{2}} \fmslash W^-
           + J_{21}^{(L, \theta)}
           + {\cal O}(\theta^2)
\, ,
\label{eq:JL21}
\\[0.2cm]
J_{11}^{(L)} &=& \left[
           e \, Q_{\psi_{\mbox{\tiny up}}} \, \fmslash A
           + \frac{g}{\cos \theta_W}
           (T_{3,\psi_{\mbox{\tiny up},L}}
           -  Q_{\psi_{\mbox{\tiny up}}}\,  \sin^2 \theta_W)
            \fmslash Z
           \right]
\nonumber \\ & &
           + J_{11}^{(L,\theta)}
 + {\cal O}(\theta^2),
\label{eq:JL11}
       \\[0.2cm]
\nonumber
J_{22}^{(L)} &=& \left[
           e \, Q_{\psi_{\mbox{\tiny down}}} \, \fmslash A
           + \frac{g}{\cos \theta_W}
           (T_{3,\psi_{\mbox{\tiny down},L}}
           -  Q_{\psi_{\mbox{\tiny down}}}\,  \sin^2 \theta_W)
            \fmslash Z
           \right]
\nonumber \\ & &
           + J_{22}^{(L,\theta)}
           + {\cal O}(\theta^2),
\label{eq:JL22}
\end{eqnarray}
\label{eq:JLcurr}
\end{subequations}
where
\begin{eqnarray}
J_{12}^{(L,\theta)} & = & \frac{g}{2\sqrt{2}} \, \theta^{\mu \nu \rho}
   \, W_{\mu}^+
 \left\{
 - \, i  \;
 \stackrel{\leftarrow}{\partial}_{\nu} \,
 \stackrel{\rightarrow}{\partial}_{\rho}
  \right.
\nonumber
  \\ & &
  \left.
 + \,  e
   \Bigg[
     Q_{\psi_{\mbox{\tiny up}}}  \:
   A_{\nu} \,  \, \stackrel{\rightarrow}{\partial}_{\rho}
 + \, Q_{\psi_{\mbox{\tiny down}}}\;  A_{\nu} \,
   \stackrel{\leftarrow}{\partial}_{\rho}
 + (Q_{\psi_{\mbox{\tiny up}}}
   +Q_{\psi_{\mbox{\tiny down}}}) \, (\partial_{\rho} A_{\nu})
   \Bigg]
  \right.
\nonumber
  \\ & &
  \left.
 + \, \frac{g}{\cos \theta_W} \,
   \Bigg[
         \bigg(
  T_{3,\psi_{\mbox{\tiny up},L}}
           -
  Q_{\psi_{\mbox{\tiny up}}} \,
      \sin^2 \theta_W
         \bigg)
   Z_{\nu} \, \stackrel{\rightarrow}{\partial}_{\rho}
  \right.
\nonumber
  \\ & &
  \left.
   \quad + \,
           (T_{3,\psi_{\mbox{\tiny down},L}}
           -  Q_{\psi_{\mbox{\tiny down}}}\,  \sin^2 \theta_W)
    \; Z_{\nu} \,\stackrel{\leftarrow}{\partial}_{\rho}
  \right.
\nonumber
  \\ & &
  \left.
   \quad + \,
       \bigg(
           (T_{3,\psi_{\mbox{\tiny up},L}}
+ T_{3,\psi_{\mbox{\tiny down},L}})
           -  (Q_{\psi_{\mbox{\tiny up}}}
+ Q_{\psi_{\mbox{\tiny down}}})
\,  \sin^2 \theta_W
       \bigg)
    \; (\partial_{\rho} Z_{\nu})
   \Bigg]
  \right.
\nonumber
  \\ & &
  \left.
- \,  \frac{i \, e \,g}{\cos \theta_W} \,
  (Q_{\psi_{\mbox{\tiny up}}}\,T_{3,\psi_{\mbox{\tiny down},L}}
  -Q_{\psi_{\mbox{\tiny down}}}\,T_{3,\psi_{\mbox{\tiny up},L}})
  \; A_{\nu}\,  Z_{\rho}
   \right\}
\label{eq:JL12theta}
\end{eqnarray}
and
\begin{eqnarray}
J_{11}^{(L,\theta)} & = & \frac{1}{2} \, \theta^{\mu \nu \rho}
 \left\{
  \, i\,  e \, Q_{\psi_{\mbox{\tiny up}}}
      \: (\partial_{\nu} A_{\mu}) \,
   \stackrel{\rightarrow}{\partial}_{\rho}
  \right.
\nonumber
  \\ & &
  \left.
 + \, \frac{i \, g}{\cos \theta_W} \,
           (T_{3,\psi_{\mbox{\tiny up},L}}
           -  Q_{\psi_{\mbox{\tiny up}}}\,  \sin^2 \theta_W)
      \: (\partial_{\nu} Z_{\mu}) \,
   \stackrel{\rightarrow}{\partial}_{\rho}
  \right.
\nonumber
  \\ & &
  \left.
 - \, e^2 \, Q_{\psi_{\mbox{\tiny up}}}^2
      \: (\partial_{\rho} A_{\mu}) \, A_{\nu}
  \right.
\nonumber
  \\ & &
  \left.
 - \, \frac{g^2}{\cos^2 \theta_W} \,
           (T_{3,\psi_{\mbox{\tiny up},L}}
           -  Q_{\psi_{\mbox{\tiny up}}}\,  \sin^2 \theta_W)^2
      \: (\partial_{\rho} Z_{\mu}) \, Z_{\nu}
  \right.
\nonumber
  \\ & &
  \left.
 - \,
       \frac{e \,g}{\cos \theta_W}
   \, Q_{\psi_{\mbox{\tiny up}}} \,
           (T_{3,\psi_{\mbox{\tiny up},L}}
           -  Q_{\psi_{\mbox{\tiny up}}}\,  \sin^2 \theta_W)
  \: \left[
       (\partial_{\rho} A_{\mu}) \, Z_{\nu}
     - A_{\mu} \, (\partial_{\rho} Z_{\nu})
     \right]
  \right.
\nonumber
  \\ & &
  \left.
 - \, \frac{g^2}{2}
  \: \left[
    W_{\mu}^+\,  W_{\nu}^- \,
   \stackrel{\rightarrow}{\partial}_{\rho}
    + (\partial_{\rho} W^+_{\mu}) \, W_{\nu}^-
     \right]
  \right.
\nonumber
  \\ & &
  \left.
 +
   \, \frac{i \, e \, g^2}{2}
   \, (2 \, Q_{\psi_{\mbox{\tiny up}}} - Q_{\psi_{\mbox{\tiny down}}})
   \; W_{\mu}^+ \, W_{\nu}^- \, A_{\rho}
  \right.
\nonumber
  \\ & &
  \left.
 + \, \frac{i \, g^3}{2 \cos \theta_W}
           \left[
        (2 \, T_{3,\psi_{\mbox{\tiny up},L}}-T_{3,\psi_{\mbox{\tiny down},L}})
       -  (2 \, Q_{\psi_{\mbox{\tiny up}}} -Q_{\psi_{\mbox{\tiny down}}})
        \,  \sin^2 \theta_W
           \right]
  \right.
\nonumber
  \\ & &
  \left.
     \times
   \; W_{\mu}^+ \, W_{\nu}^- \, Z_{\rho}
  \right\}
\, ,
\label{eq:JL11theta}
\end{eqnarray}
while
\begin{equation}
\left.
\begin{array}{l}
J_{21}^{(L,\theta)} \\[0.2cm]
J_{22}^{(L,\theta)}
\end{array}
\right\}
 =
\left\{
\begin{array}{l}
J_{12}^{(L,\theta)} \\[0.2cm]
J_{11}^{(L,\theta)}
\end{array}
\right.
(
W^+  \leftrightarrow  W^- ,
 Q_{\psi_{\mbox{\tiny up}}}
 \leftrightarrow  Q_{\psi_{\mbox{\tiny down}}} ,
 T_{3,\psi_{\mbox{\tiny up},L}}
 \leftrightarrow  T_{3,\psi_{\mbox{\tiny down},L}}
)\, .
\label{eq:JL21thetaJL22theta}
\end{equation}
Here and in the following we use the notation in which
$\stackrel{\rightarrow}{\partial}_{\rho}$
denotes the partial derivative which acts {\it only} on the fermion field
on the right side, while
$\stackrel{\leftarrow}{\partial}_{\rho}$
denotes the partial derivative which acts {\it only} on the fermion field
on the left side, i.e.
\begin{equation}
{\partial}_{\rho} \psi \equiv
\stackrel{\rightarrow}{\partial}_{\rho} \psi
 \qquad
{\partial}_{\rho} \overline{\psi} \equiv
\overline{\psi} \stackrel{\leftarrow}{\partial}_{\rho}
\, .
\label{eq:partial}
\end{equation}
We note that in contrast to the SM case, although
\begin{displaymath}
\left( \int d^4\! x\,
 \bar\psi_{\mbox{\tiny up},L} \, J_{12} \, \psi_{\mbox{\tiny down},L}
\right)^{\dagger} =
 \int d^4\! x\,
 \bar\psi_{\mbox{\tiny down},L} \, J_{21} \, \psi_{\mbox{\tiny up},L}
\, ,
\end{displaymath}
we have
\begin{displaymath}
J_{21}^{(L)}
\ne \gamma^0 \,
\left( J_{12}^{(L)} \right) ^{\dagger}
\, \gamma_0
\, .
\end{displaymath}
The reason is the specific form of the interaction
term (see Eq.~\req{eq:hatPsiDPsi}) which contains derivatives,
whose pressence produce
\begin{displaymath}
J_{21}^{(L)}
= \gamma^0 \,
\left( J_{12}^{(L)}
(\stackrel{\rightarrow}{\partial}
\leftrightarrow
\stackrel{\leftarrow}{\partial}
)\right) ^{\dagger}
\, \gamma_0
\,.
\end{displaymath}

Now, we turn to the results for the right-handed electroweak sector.
Here $\psi_R$ represents $\psi_R \in \{ e_R^{(i)}, u_R^{(i)}, d_R^{(i)} \}$,
and the representation ${\newrho}_{\psi_R}(V_{\mu})$
from Table  \ref{tab:rhoPsi} without $SU(3)$ fields is given by
\begin{equation}
{\newrho}_{\psi_R}(V_{\mu})=
g' \, {\cal A}_{\mu} \, Y_{\psi_R}
=e \, Q_{\psi} \, A_{\mu}
- \frac{g}{\cos \theta_W} \, Q_{\psi} \sin^2 \theta_W \, Z_{\mu}
\,.
\label{eq:right}
\end{equation}
For the right-handed fermions,
$T_{3,\psi_R}=0$ and $Y_{\psi_R}=Q_{\psi}$.
The right-handed electroweak part
of the action $S_{\mbox{\tiny $\psi$}}$ 
is of the form
\begin{eqnarray}
S_{\mbox{\tiny $\psi$,ew,R}} &=&
 \int d^4 x \left( \bar\psi_R\, i\fmslash \partial\, \psi_R +
\bar\psi_R \, J^{(R)} \, \psi_R \right)
\, ,
\label{eq:SewR}\\
J^{(R)} &=& \left[
           e \, Q_{\psi} \, \fmslash A
           - \frac{g}{\cos \theta_W}
           Q_{\psi} \,  \sin^2 \theta_W
            \fmslash Z
           \right]
           + J^{(R,\theta)}
           + {\cal O}(\theta^2)
\, ,
\label{eq:JRcurr}\\
J^{(R,\theta)} & = & \frac{1}{2} \, \theta^{\mu \nu \rho}
  \left\{
 \, i \, e \, Q_{\psi} \,
  \: (\partial_{\nu} A_{\mu}) \,
\stackrel{\rightarrow}{\partial}_{\rho}
- \, \frac{i \, g}{\cos \theta_W} \, Q_{\psi} \, \sin^2 \theta_W
  \: (\partial_{\nu} Z_{\mu}) \,
\stackrel{\rightarrow}{\partial}_{\rho}
\right.
\nonumber
\\ & &
\left.
- \, e^2 \, Q_{\psi}^2
  \: (\partial_{\rho} A_{\mu}) \, A_{\nu}
- \, \frac{g^2}{\cos^2 \theta_W} \, Q_{\psi}^2 \, \sin^4 \theta_W
  \: (\partial_{\rho} Z_{\mu}) \, Z_{\nu}
\right.
\nonumber
\\ & &
\left.
+
  \, \frac{e \,g}{\cos \theta_W}
  \,  Q_{\psi}^2 \, \sin^2 \theta_W
  \: \left[
   (\partial_{\rho} A_{\mu}) \, Z_{\nu}
   -  \, A_{\mu} \, (\partial_{\rho} Z_{\nu})
     \right]
\right\}
\, .
\label{eq:JRtheta}
\end{eqnarray}

Let us now present our results
\footnote{We note that in the preceding considerations
we have corrected the electroweak currents
presented in the appendix of Ref. \cite{Calmet:2001na}
and expressed them using more compact and transparent notation.}
in a form
suitable for further calculations, derivation
of Feynman rules and phenomenological
applications, i.e. in terms of 
$\Psi \in \{ L^{(i)}, Q^{(i)} \}$,
and thus
$\psi_{\mbox{\tiny up}} \in \{ \nu^{(i)}, u^{(i)}  \}$,
and
$\psi_{\mbox{\tiny down}} \in \{ e^{(i)}, d^{(i)} \}$.
The electroweak part of the action
$S_{\mbox{\tiny $\psi$}}$
then takes the form
\begin{eqnarray}
S_{\mbox{\tiny $\psi$,ew}}
 &=&
 \int d^4 x \Bigg\{ \bar\Psi\, i\fmslash \partial\, \Psi
\nonumber \\
 &&
\left.
+ \,
\bar\psi_{\mbox{\tiny up}} \, J_{12}^{(L)} \,
 \, \frac{1}{2} \, (1-\gamma_5) \, \psi_{\mbox{\tiny down}}
+ \,
\bar\psi_{\mbox{\tiny down}} \, J_{21}^{(L)} \,
\, \frac{1}{2} \, (1-\gamma_5) \,\psi_{\mbox{\tiny up}}
\right.
\nonumber \\
 &&
\left.
\, + \,
\bar\psi_{\mbox{\tiny up}}
\; \frac{1}{2} \left[
   (J_{11}^{(L)}+J^{(R)})
   - (J_{11}^{(L)}-J^{(R)}) \gamma_5
               \right]
\, \psi_{\mbox{\tiny up}}
\right.
\nonumber \\
 &&
\left.
+ \,
\bar\psi_{\mbox{\tiny down}}
\; \frac{1}{2} \left[
   (J_{22}^{(L)}+J^{(R)})
   - (J_{22}^{(L)}-J^{(R)}) \gamma_5
               \right]
\, \psi_{\mbox{\tiny down}}
\right\}
\, ,
\label{eq:Sew}
\end{eqnarray}
and the currents $J_{ij}^{(L)}$ can be read from
Eqs. (\ref{eq:JLcurr}-\ref{eq:JL21thetaJL22theta}),
while $J^{(R)}$ is given by Eqs. (\ref{eq:JRcurr}-\ref{eq:JRtheta})
(with $Q_{\psi}$ substituted by the corresponding
$Q_{\psi_{\mbox{\tiny up}}}$ or
$Q_{\psi_{\mbox{\tiny down}}}$).

Finally, we note that the fermion fields appearing in this section
are not mass but weak-interaction eigenstates. In order to present
the results in terms of mass eigenstates, the
Cabbibo-Kobayashi-Maskawa matrix (denoted by $V_{ij}$ in the
following) enters the quark currents leading to mixing between
generations and to the modification of the quark currents by
$V_{ij}$ factors:
\begin{displaymath}
\bar{q}_{\mbox{\tiny up}}^{(i)}
\, V_{ij} \, J_{12}^{(L)} \,
\, \frac{1}{2} \, (1-\gamma_5)
\,
q_{\mbox{\tiny down}}^{(j)} \, ,
\qquad
\bar{q}_{\mbox{\tiny down}}^{(j)}
\, V_{ij}^* \, J_{21}^{(L)} \,
\, \frac{1}{2} \, (1-\gamma_5)
\,
q_{\mbox{\tiny up}}^{(i)} \, ,
\end{displaymath}
where $q_{\mbox{\tiny up}}^{(i)}$ and $q_{\mbox{\tiny down}}^{(i)}$
represent mass eigenstates. In the NCSM, as in the SM, the neutrino
masses are not considered and consequently the leptonic mixing
matrix is diagonal in contrast to the neutrino mass extended models.
The corresponding non-commutative extensions which include neutrino
masses can be made along the lines sketched here (see Section
\ref{Y} for further details on this subject).

In this section, only electroweak interactions were
considered. Pure QCD, as well as mixed terms which appear in the
NCSM due to the Seiberg-Witten mapping, are left for a future
publication \cite{Melic:2005am}.

\section{\label{sec:higgs}Higgs Sector of the NCSM Action}

In the preceding section we have expanded the fermionic part of the action
and performed a detailed analysis of the electroweak interactions.
We devote this section to the
analysis of $S_{\mbox{\tiny Higgs}}$ and $S_{\mbox{\tiny Yukawa}}$
to first order in $\theta$.

\subsection{\label{H}Higgs Kinetic Terms}

The expansion of the Higgs part of the action (\ref{eq:Shiggs})
to first order in $\theta$ yields
\footnote{In order to make the presentation more
transparent, in this section,
we denote the $2 \times 2$ matrices appearing in the action by bold letters.}
\begin{eqnarray}
S_{\mbox{\tiny{Higgs}}} {\hspace{-1mm}}&=&
 \int d^4x \bigg( ({\bf D}_\mu \Phi)^\dagger
 ({\bf D}^\mu \Phi)
- \mu^2 {\Phi}^\dagger  \Phi
- \lambda \,
(\Phi^\dagger  \Phi)^2
\bigg)
\nonumber \\ &&
+ \frac{1}{2} \, \theta^{\alpha\beta}\int {\hspace{-1mm}}d^4 x\,
\Phi^\dagger \Bigg( {\bf U}_{\alpha
    \beta} + {\bf U}^\dagger_{\alpha \beta}
    + \frac12  \mu^2 \, {\bf F}_{\alpha\beta} -
    2 i\lambda\, \Phi ({\bf D}_\alpha\Phi)^\dagger
    {\bf D}_\beta \Bigg)\Phi \,,
\nonumber \\ &&
\label{eq:H}
\end{eqnarray}
where
\begin{eqnarray}
{\bf U} _{\alpha \beta} &=& \left(
    \stackrel{\leftarrow}{{\bf \pat}^\mu}
    + \, i{\bf V}^\mu \right) \Bigg(
    -\, {\bf \pat}_\mu {\bf V}_\alpha \, {\bf \pat}_\beta
    - \, {\bf V}_\alpha {\bf \pat}_\mu{\bf \pat}_\beta
    + \, {\bf \pat}_\alpha {\bf V}_\mu {\bf \pat}_\beta
 \nonumber \\ & &
    + \, i {\bf V}_\mu {\bf V}_\alpha {\bf \pat}_\beta
    + \, \frac{i}2 {\bf V}_{\alpha}{\bf V}_{\beta}{\bf \pat}_{\mu}
+ \, \frac{i}2 {\bf \pat}_{\mu}({\bf V}_{\alpha}{\bf V}_{\beta})
  \nonumber \\ & &
+ \, \frac12 {\bf V}_{\mu}{\bf V}_{\alpha}{\bf V}_{\beta}
+ \, \frac{i}2 \{{\bf V}_\alpha,\, {\bf \pat}_\beta {\bf V}_\mu + {\bf F}_{\beta\mu} \}
\Bigg).
\label{aa}
\end{eqnarray}
Equation \req{eq:H} contains
the usual covariant derivative of the Higgs boson
${\bf D}_\mu=  \partial_{\mu} {\bf 1}-i {\bf V}_{\mu}$ where
${\bf V}_{\mu}=g' {\mathcal A}_\mu Y_{\Phi} {\bf 1} + g B_\mu^a T_L^a$,
and ${\bf 1}$ is a unit matrix
suppressed in the following. 
Also
${\bf F}_{\mu \nu}=
\partial_{\mu} {\bf V}_{\nu}-\partial_{\nu} {\bf V}_{\mu}
 -i [{\bf V}_{\mu},{\bf V}_{\nu}]$\,.

Let us construct explicit expressions for the
electroweak gauge matrices occurring
in (\ref{eq:H}) and (\ref{aa}).
The gauge field ${\bf V}_\mu$
can be expressed in a matrix form as
\begin{equation}
{\bf V}_\mu = \left( \begin{array}{cc}
\displaystyle
g'\mathcal A_\mu Y_{\Phi} + g T_{3,\phi_{\mbox{\tiny up}}}
    B_\mu^3
    &
\displaystyle
\frac{g}{\sqrt{2}} W_{\mu}^+\\[0.4cm]
\displaystyle
    \frac{g}{\sqrt{2}} W_{\mu}^- &
\displaystyle
g'\mathcal A_\mu Y_{\Phi} + g T_{3,\phi_{\mbox{\tiny down}}}
    B_\mu^3 \end{array} \right)
\, ,
\label{eq:V}
\end{equation}
where from Table \ref{tab:SMfields} one can read%
\footnote{Note
$
Y_{\Phi}=Q_{\phi_{\mbox{\tiny up}}}-T_{3,\phi_{\mbox{\tiny up}}}
        =Q_{\phi_{\mbox{\tiny down}}}-T_{3,\phi_{\mbox{\tiny down}}}
\, .
$
}:
$Y_{\Phi}=1/2$, $T_{3,\phi_{\mbox{\tiny up}}}=1/2$, $T_{3,\phi_{\mbox{\tiny
down}}}=-1/2$.
The diagonal matrix elements can also be expressed in terms of physical
fields using Eqs. (\ref{eq:Amu}).
Hence, one obtains
\begin{eqnarray}
V_{11,\mu} &=& e A_{\mu}
   + \frac{g}{2 \cos \theta_W} (1-2 \sin^2 \theta_W) Z_{\mu}
\, ,
\nonumber \\
V_{22,\mu} &=& - \frac{g}{2 \cos \theta_W} Z_{\mu}
\, .
\label{eq:V1122}
\end{eqnarray}
The product of two gauge fields is given by
\begin{equation}
{\bf V}_\mu {\bf V}_\alpha =  \left( \begin{array}{cc}
\displaystyle
    V_{11,\mu} V_{11,\alpha} + \frac{g^ 2}2W_\mu^+
    W_\alpha^-&
\displaystyle
    \frac{g}{\sqrt{2}} \left( W_\alpha^+ V_{11,\mu} +
    W_\mu^+ V_{22,\alpha} \right)\\[0.4cm]
\displaystyle
    \frac{g}{\sqrt{2}} \left( W_\alpha^- V_{22,\mu} +
    W_\mu^- V_{11,\alpha} \right)&
\displaystyle
    V_{22,\mu} V_{22,\alpha} +
    \frac{g^2}{2}W_\mu^- W_\alpha^+
    \end{array}\right),
\label{eq:VV}
\end{equation}
while the product of three gauge fields can be expressed as
\begin{subequations}
\label{eq:VVVM}
\begin{equation}
\label{eq:VVV}
{\bf V}_\mu {\bf V}_\alpha {\bf V}_\beta =  \bf M_{\mu \alpha \beta},
\end{equation}
with matrix elements
\begin{eqnarray}
\nonumber
M_{\mu \alpha \beta,11} & = &
   V_{11,\mu} V_{11,\alpha} V_{11,\beta}
    \\
\nonumber
&& + \frac{g^2}2 (V_{11,\mu}W^+_\alpha W^-_\beta
        + W^+_\mu W^-_\alpha V_{11,\beta}
        + W^+_\mu V_{22,\alpha}W^-_\beta) \, ,\\
\nonumber
M_{\mu \alpha \beta,12} & = & \frac{g}{\sqrt{2}} \Big(V_{11,\mu}
    W_\alpha^+ V_{22,\beta} + V_{11,\mu}
    V_{11,\alpha} W_\beta^+ + W_\mu^+ V_{22,\alpha}
    V_{22,\beta}
\\
\nonumber
&& + \frac{g^2}{2} W^+_\mu W^-_\alpha W^+_\beta \Big)\, ,\\
\nonumber
M_{\mu \alpha \beta,21} & = & \frac{g}{\sqrt{2}} \Big(
    V_{22,\mu} W^-_\alpha V_{11,\beta}
    +V_{22,\mu} V_{22,\alpha}W^-_\beta
    +W^-_\mu V_{11,\alpha} V_{11,\beta}\\
\nonumber
&& + \frac{g^2}{2} W^-_\mu W^+_\alpha W^-_\beta \Big)\, ,\\
\nonumber
M_{\mu \alpha \beta,22} & = & V_{22,\mu} V_{22,\alpha}
    V_{22,\beta}\\
&& + \frac{g^2}2 (V_{22,\mu}W^-_\alpha W^+_\beta
        + W^-_\mu W^+_\alpha V_{22,\beta}
        + W^-_\mu V_{11,\alpha}W^+_\beta) \, .
\label{eq:M}
\end{eqnarray}
\end{subequations}
For the field strength one obtains
\begin{eqnarray}
\nonumber
{\bf F}_{\mu\nu}
& = & \left(
\begin{array}{cc}
\displaystyle
 e A_{\mu \nu}
   + \frac{g}{2 \cos \theta_W} (1-2 \sin^2 \theta_W) Z_{\mu \nu}
&
\displaystyle
\frac{g}{\sqrt{2}} W^+_{\mu\nu} \\[0.4cm]
\displaystyle
\frac{g}{\sqrt{2}} W^-_{\mu\nu} &
\displaystyle
 - \frac{g}{2 \cos \theta_W} Z_{\mu \nu}
\end{array}\right)
\nonumber \\[0.4cm] &&
- \frac{ig^2}2\,
\left(
\begin{array}{cc}
\displaystyle
W^+_\mu W^-_\nu - W^+_\nu W^-_\mu &
\displaystyle
\sqrt{2} (B^3_\mu W^+_\nu - W_\mu^+ B_\nu^3) \\[0.2cm]
\displaystyle
-\sqrt{2} (B^3_\mu W^-_\nu - W_\mu^- B_\nu^3) &
\displaystyle
- W^+_\mu W^-_\nu + W^+_\nu W^-_\mu
\end{array}\right)
\, , \qquad
\label{vmunu}
\end{eqnarray}
where
$X_{\mu\nu}=\pat_\mu  X_\nu - \pat_\nu X_\mu$
for $X \in \{A,Z,W^+,W^-\}$.
By making use of Eq. \req{eq:Amu} one can completely express
the off-diagonal elements
in terms of the physical fields $A_\mu$ and $Z_\mu$.
The other combinations of fields appearing in Eqs.~(\ref{eq:H})
and (\ref{aa}) can also
be easily obtained. We will not provide the explicit expressions here.

It is not difficult to see that the value of the Higgs field that
minimizes the (non-commutative) Higgs potential is the same as in
the commutative case because of the following:
We are looking for the minimum value of the
potential attained for \emph{constant} fields and hence can ignore
all derivative terms and all star products. This leaves terms like
$\theta^{\alpha\beta} V_\alpha V_\beta \Phi$ in the hybrid SW map
that could possibly lead to corrections of the vacuum expectation
value of the Higgs. Taking into account also the potential of the gauge fields
it is, however, clear that we should consider only
$V_\alpha =0$, i.e.\ $\widehat\Phi = \Phi$ when fixing the vacuum
expectation value.

The Higgs field is chosen to be in the unitary gauge
\begin{equation}
\Phi(x) \equiv \phi(x) = \frac1{\sqrt{2}} \left(\begin{array}{c}0\\ h(x) 
+ v\end{array}\right)
\, ,
\label{eq:Higgs}
\end{equation}
where $v=\sqrt{-\mu^2/\lambda}$
represents the Higgs vacuum expectation value, while $h(x)$ 
is the physical Higgs field.

There are several points that need to be mentioned in connection
with the NCSM version of the $S_{\mbox{\tiny Higgs}}$ part of the
action \req{eq:H}. From \req{eq:Higgs} one trivially obtains
$$
\int d^4x \, \phi^\dagger H \phi
= \int d^4x (h(x)+v) \, H_{22} \, (h(x)+v)
\, ,
$$
where $H$ stands here for any $2\times 2$
matrix. Taking into account this along with \req{eq:V} and
(\ref{eq:VV}-\ref{vmunu}), it is easy to see that terms
containing one or more Higgs fields $h(x)$ as well as terms
containing solely gauge bosons reside in \req{eq:H}.

First, let us examine the contributions of the last two
$\theta$-dependent terms in Eq.  (\ref{eq:H}). By making use of
(\ref{eq:H}-\ref{vmunu}) for the Higgs field in unitary gauge we
find
\begin{eqnarray}
\lefteqn{
\frac{1}{2}\theta^{\alpha\beta}\int {\hspace{-1mm}}d^4 x\,
\phi^\dagger \Bigg(\frac12  \mu^2 \, {\bf F}_{\alpha\beta} -
    2 i\lambda\, \phi ({\bf D}_\alpha\phi)^\dagger
    {\bf D}_\beta \Bigg)\phi
}
    &&
    \nonumber \\
&=&\frac{1}{8} \theta^{\alpha \beta}
\left\{
i g^2
\int d^4 x
(h+v)^2 [\mu^2 + \lambda (h+v)^2 ]
W^+_{\alpha} W^-_{\beta}
\right.
\nonumber \\ & &
\left.
+ \frac{g}{\cos \theta_W}
\int d^4 x
(h+v)^2
[- \mu^2 (\pat_{\alpha} Z_{\beta})
+ 2 \lambda (h+v) (\pat_{\alpha} h) Z_{\beta}]
\right\}
\, . \qquad
\label{eq:higgsLambda}
\end{eqnarray}
Owing to the Stokes theorem
the term containing only one $Z$ field
vanishes. Similarly, by performing partial integration
and taking into account $v^2=-\mu^2/\lambda$,
the spuriously looking two-field terms vanish and
\req{eq:higgsLambda}
simplifies to
\begin{equation}
\frac{1}{8} \theta^{\alpha \beta}
\lambda
\int d^4 x \,
h (h+v)(h+ 2 v)
\left\{
i g^2
 (h+v)
W^+_{\alpha} W^-_{\beta}
+ 2 \frac{g}{\cos \theta_W}
 (\pat_{\alpha} h) Z_{\beta}
\right\}
\, .
\label{eq:higgsLambdaFin}
\end{equation}

Second, let us note that, in contrast to the SM case, in the NCSM
action $S_{\mbox{\tiny Higgs}}$
\req{eq:H} there are terms proportional to $v^2$ that cannot
be identified  as the mass terms of the Higgs and weak gauge bosons
fields but represent interaction terms.
Hence, after the identification
of the mass terms ($- 1/2 \, m_H^2 h^2$),
$M_W^2 W_{\mu}^+ W^{- \, \mu}$ and
$1/2 M_Z^2 \, Z_{\mu} Z^{\mu} $
with Higgs, W and Z boson masses
\begin{eqnarray}
& & m_H^2=2 \mu^2=- 2 v^2 \lambda
\, , \nonumber \\
& &M_W^2 =\frac{1}{4} v^2 g^2
\, ,
\qquad
M_Z^2 =\frac{1}{4} v^2 (g^2+{g'}^2)
=\frac{M_W^2}{\cos^2 \theta_W}
\, ,
\end{eqnarray}
respectively, additional terms remain which describe interactions of
Higgs and gauge bosons and interactions of solely gauge bosons. The
latter behaviour is novel in comparison with the Standard Model and
is introduced by the Seiberg-Witten mapping. The analysis of Eq.
\req{eq:H} reveals that, in addition to the interaction terms
contained in $S_{\mbox{\tiny gauge}}$ \req{eq:genSgaugeSM}, the last
three terms of the second bracket in ${\bf U} _{\alpha \beta}$
\req{aa} give rise to order $\theta$ contributions to the three- and
four-gauge-boson couplings. Specifically, the three-gauge-boson
interaction terms from $S_{\mbox{\tiny Higgs}}$ read $(-1/4)v^2 \,
\theta^{\alpha \beta} [I_{\alpha \beta} + I^{\dagger}_{\alpha
\beta}]_{22}\,, $ where $ I_{\alpha \beta} = {\bf V}^{\mu}
 [ (\partial_{\mu} {\bf V}_{\alpha}) {\bf V}_{\beta}
   + {\bf V}_{\alpha} (\partial_{\beta} {\bf V}_{\mu})
   + (\partial_{\beta} {\bf V}_{\mu}) {\bf V}_{\alpha} ]
\, .
$
By making use of \req{eq:VVVM} one arrives at explicit expressions for
the  $W^+ W^- \gamma$, $W^+ W^- Z$ and $ZZZ$ interaction terms:
\begin{eqnarray}
\lefteqn{\frac{-1}{4} \, v^2 \, \theta^{\alpha \beta}
\left[ I_{\alpha \beta} + I^{\dagger}_{\alpha \beta}
\right]_{22}}
\nonumber \\[0.2cm]
&=& \frac{e}{2} \, M_W^2 \, \theta^{\alpha \beta}
\left[
\left(
W^{+\mu} \, W_{\alpha}^- +
W^{-\mu} \, W_{\alpha}^+ \right)
A_{\mu \beta}
+ \left( \partial_{\beta} A_{\alpha} \right)
W^{+\mu} \, W^-_{\mu} \right]
\nonumber \\[0.2cm]
&& - \frac{g}{4 \cos \theta_W} M_W^2 \, \theta^{\alpha \beta}
\left\{
 Z^{\mu}
\left[ W_{\mu}^+ \, \left(\partial_{\beta} W^-_{\alpha}\right) +
       W_{\mu}^- \, \left(\partial_{\beta} W^+_{\alpha}\right) \right]
\right.
\nonumber \\[0.2cm]
&&
\left.
\hspace*{0.2cm}
+ \left( Z^{\mu} W_{\alpha}^+
+ Z_{\alpha} W^{+\mu} \right)
W_{\mu \beta}^-
+ \left( Z^{\mu} W_{\alpha}^-
+ Z_{\alpha} W^{-\mu} \right)
W_{\mu \beta}^+
\right.
\nonumber \\[0.2cm]
&&
\left.
\hspace*{0.2cm}
- \cos 2 \theta_W \left[
\left(
W^{+\mu}\,  W_{\alpha}^- +
W^{-\mu}\,  W_{\alpha}^+ \right)
Z_{\mu \beta}
+ \left( \partial_{\beta} Z_{\alpha} \right)
W^{+\mu} \, W^-_{\mu} \right]
\right\}
\nonumber \\[0.2cm]
&& + \frac{g}{4 \cos \theta_W} \, M_Z^2 \, \theta^{\alpha \beta}
\, Z^{\mu}\,  Z_{\alpha}
\left(
2 \, \partial_{\beta} Z_{\mu}
- \partial_{\mu} Z_{\beta}
\right)\, .
\nonumber \\
\label{eq:higgsWWA}
\end{eqnarray}
The four-gauge-boson interaction
terms can be analysed analogously.

\subsection{\label{Y}Yukawa Terms}

Next, we proceed to the $\theta$-expansion
of the $S_{\mbox{\tiny Yukawa}}$
action \req{eq:Syukawa}.
Similarly to the analysis of the electroweak currents
presented in Section \ref{sec:ew},
let us first analyse
the general form for the Yukawa action,
\begin{eqnarray}
S_{\mbox{\tiny $\psi$, Yukawa}} & = &
 - \int d^4x \sum_{i,j=1}^3
\left[
\left(
G_{\mbox{\tiny down}}^{(ij)} \,
( \overline{ \widehat \Psi}^{(i)}_L  *  h_{\psi_{\mbox{\tiny down}}}(\widehat \Phi)
 * \widehat \psi_{\mbox{\tiny down},R}^{(j)} )
+ \mbox{h.c.}
\right)
\right.
\nonumber
\\ & &
\left.
+
\left(
G_{\mbox{\tiny up}}^{(ij)}
( \overline{ \widehat \Psi}^{(i)}_L  *  h_{\psi_{\mbox{\tiny up}}}(\widehat \Phi_c)
 * \widehat \psi_{\mbox{\tiny up},R}^{(j)} )
+ \mbox{h.c.}
\right)
\right]
\, .
\label{eq:Syukawa-gen}
\end{eqnarray}
Here $G_{\mbox{\tiny down}}$ and $G_{\mbox{\tiny up}}$ are general
$3\times 3$ matrices which comprise Yukawa couplings while
$\psi_{\mbox{\tiny up},R}^{(j)}$ and $\psi_{\mbox{\tiny
down},R}^{(j)}$ denote up and down fermion fields of the generation
$j$. As we analyse a simple non-commutative extension of the SM,
$G_{\mbox{\tiny up}}^{ij}$ vanishes for leptons. Furthermore, as in
the SM one can find a biunitary transformation that diagonalizes the
$G$ matrices
\begin{eqnarray*}
G_{\mbox{\tiny down}} &=& \frac{\sqrt{2}}{
\, v}
 \, S_{\mbox{\tiny down}}
 \, M_{\mbox{\tiny down}}
 \, T_{\mbox{\tiny down}}^\dagger \,,
\;\;\;
G_{\mbox{\tiny up}} = \frac{\sqrt{2}}{
\, v}
 \, S_{\mbox{\tiny up}}
 \, M_{\mbox{\tiny up}}
 \, T_{\mbox{\tiny up}}^\dagger\,,
\end{eqnarray*}
and obtain the diagonal $3\times 3$ mass matrices
$M_{\mbox{\tiny down}}$
and
$M_{\mbox{\tiny up}}$.
Next, one redefines the fermion fields to mass eigenstates
\begin{eqnarray*}
\overline{\widehat \psi}_{\mbox{\tiny down},L}^{(i)} S_{\mbox{\tiny down}}^{(ij)}
\rightarrow
\overline{\widehat \psi}_{\mbox{\tiny down},L}^{(j)}
&\quad&
 \, T_{\mbox{\tiny down}}^{\dagger(ij)}
\widehat \psi_{\mbox{\tiny down},R}^{(j)}
\rightarrow
\widehat \psi_{\mbox{\tiny down},R}^{(i)}
\nonumber \\
\overline{\widehat \psi}_{\mbox{\tiny up},L}^{(i)} S_{\mbox{\tiny up}}^{(ij)}
\rightarrow
\overline{\widehat \psi}_{\mbox{\tiny up},L}^{(j)}
&\quad&
 \, T_{\mbox{\tiny up}}^{\dagger(ij)}
\widehat \psi_{\mbox{\tiny up},R}^{(j)}
\rightarrow
\widehat \psi_{\mbox{\tiny up},R}^{(i)}
\, .
\end{eqnarray*}
This redefinition of the fields introduces the
fermion mixing matrix
$V = S_{\mbox{\tiny up}}^\dagger
 \, S^{}_{\mbox{\tiny down}}$ \,
in the electroweak currents \req{eq:Sew}, and, owing to the hybrid
SW mapping of the Higgs field, in the Yukawa part of the NCSM action
as well. We introduce the matrix $V_f$, which like in the SM,
corresponds to
\begin{equation}
  V_f = \left\{
\begin{array}{cc}
{\mathbf 1}
&
\mbox{for } f= \ell
\\
V \equiv V_{CKM}
&
\mbox{for } f=q
\end{array}
\right.
\, ,
\label{eq:fCKM}
\end{equation}
where $\ell$ and $q$ denote leptons and quarks, respectively. Hence,
the quark mixing is described by the CKM matrix, while the mixing in
the lepton sector is absent but can be additionally  introduced
following the commonly accepted modifications of the SM which
comprise neutrino masses. Furthermore, as the Higgs part of the NCSM
action introduces mass dependent gauge boson couplings (see Eq.
(\ref{eq:higgsWWA})), the Yukawa part of the NCSM action introduces
fermion mass dependent interactions. In contrast to the NCSM, in the
SM fermion mass dependent interactions always include an interaction
with the Higgs field.

Using Eq. \req{eq:def-h} we find
\begin{eqnarray}
\lefteqn{\int d^4 x
\, \overline{ \widehat \Psi}^{(i)}_L  *  h_{\psi_{\mbox{\tiny down}}}(\widehat \Phi)
 * \widehat \psi_{\mbox{\tiny down},R}^{(j)} }
\nonumber \\[0.1cm]
& = & \int d^4 x\,
( \overline{ \Psi}^{(i)}_L   \Phi
\, \psi_{\mbox{\tiny down},R}^{(j)} )
+ \frac{1}{2} \int d^4 x \,\theta^{\mu \nu}
\overline{ \Psi}^{(i)}_L
\left[
-i
\stackrel{\leftarrow}{\partial}_{\mu}
\Phi
\stackrel{\rightarrow}{\partial}_{\nu}
\right.
\nonumber \\[0.1cm]
& & \left.
-
\stackrel{\leftarrow}{\partial}_{\nu}
\,
\newrho_{\Psi_L}(V_{\mu})\,
\Phi
-
\Phi \,
\newrho_{\psi_{\mbox{\tiny down},R}}(V_{\mu})
\,
\stackrel{\rightarrow}{\partial}_{\nu}
\right.
\nonumber \\[0.1cm]
& & \left.
-
\newrho_{\Psi_L}(V_{\mu})
\,
(\partial_\mu \Phi)
-
(\partial_\mu \Phi)
\,
\newrho_{\psi_{\mbox{\tiny down},R}}(V_{\mu})
\right.
\nonumber \\[0.1cm]
& & \left.
+ i \,
\newrho_{\Psi_L}(V_{\mu}) \,
\newrho_{\Psi_L}(V_{\nu}) \,
 \Phi
+ i
 \Phi \,
\newrho_{\psi_{\mbox{\tiny down},R}}(V_{\mu}) \,
\newrho_{\psi_{\mbox{\tiny down},R}}(V_{\nu})
\right.
\nonumber \\[0.1cm]
& & \left.
- i
\newrho_{\Psi_L}(V_{\mu}) \,
 \Phi \,
\newrho_{\psi_{\mbox{\tiny down},R}}(V_{\nu})
\right.
\Big]
\psi_{\mbox{\tiny down},R}^{(j)}
\, .
\label{eq:Sy}
\end{eqnarray}
The representations
$\newrho_{\Psi_L}(V_{\mu})$ and
$\newrho_{\psi_{\mbox{\tiny down},R}}(V_{\mu})$
can be read from Table \ref{tab:rhoPsi}.
Expressions valid for both leptons and quarks,
with strong interactions omitted, are given in
Eqs. \req{eq:left} and \req{eq:right}.
For the Higgs field \req{eq:Higgs} is used.

Finally, using \req{eq:Sy}, after some algebra
we obtain the following result for
\req{eq:Syukawa-gen} expressed
in terms of physical fields
(and with gluons omitted):
\begin{eqnarray}
S_{\mbox{\tiny $\psi$, Yukawa}} & = &
\int d^4 x \sum_{i,j=1}^3
\left [
\bar{\psi}_{\mbox{\tiny down}}^{(i)}
\left(
N_{dd}^{V(ij)}
+ \gamma_5 \,
N_{dd}^{A(ij)}
\right)
\psi_{\mbox{\tiny down}}^{(j)}
\right.
\nonumber \\[0.1cm] & & \left.
+ \bar{\psi}_{\mbox{\tiny up}}^{(i)}
\left(
N_{uu}^{V(ij)}
+ \gamma_5 \,
N_{uu}^{A(ij)}
\right)
\psi_{\mbox{\tiny up}}^{(j)}
\right.
\nonumber \\[0.1cm] & & \left.
+ \bar{\psi}_{\mbox{\tiny up}}^{(i)}
\left(
C_{ud}^{V(ij)}
+ \gamma_5 \,
C_{ud}^{A(ij)}
\right)
\psi_{\mbox{\tiny down}}^{(j)}
\right.
\nonumber \\[0.1cm] & & \left.
+ \bar{\psi}_{\mbox{\tiny down}}^{(i)}
\left(
C_{du}^{V(ij)}
+ \gamma_5 \,
C_{du}^{A(ij)}
\right)
\psi_{\mbox{\tiny up}}^{(j)}
\right]
\, .
\label{eq:SYukawa-fiz}
\end{eqnarray}
The neutral currents read
\begin{eqnarray}
N_{dd}^{V(ij)}
&=&
-M_{\mbox{\tiny down}}^{(ij)} \left(1 + \frac{h}{v}\right)
+
N_{dd}^{V,\theta (ij)}
+
{\cal O}(\theta^2)
\, ,
\nonumber \\
N_{dd}^{A(ij)}
&=&
N_{dd}^{A,\theta (ij)}
+
{\cal O}(\theta^2)
\, ,
\nonumber
\\
N_{uu}^{V(ij)}
&=&
-M_{\mbox{\tiny up}}^{(ij)} \left(1 + \frac{h}{v}\right)
+
N_{uu}^{V,\theta (ij)}
+
{\cal O}(\theta^2)
\, ,
\nonumber \\
N_{uu}^{A(ij)}
&=&
N_{uu}^{A,\theta (ij)}
+
{\cal O}(\theta^2)
\, ,
\label{eq:dV}
\end{eqnarray}
where
\begin{eqnarray}
\lefteqn{N_{dd}^{V,\theta (ij)}}
\nonumber \\
&=&
- \frac{1}{2} \theta^{\mu \nu}
M_{\mbox{\tiny down}}^{(ij)}
\left\{
i \frac{(\partial_{\mu}h)}{v} \stackrel{\rightarrow}{\partial}_{\nu}
\right.
\nonumber \\ & &
\left.
-
\left[
e Q_{\psi_{\mbox{\tiny down}}} A_{\mu}
+
\frac{g}{2 \cos \theta_W}
(T_{3,\psi_{\mbox{\tiny down},L}}-
2 Q_{\psi_{\mbox{\tiny down}}} \sin^2 \theta_W) Z_{\mu}
\right]
 \frac{(\partial_{\nu}h)}{v}
\right.
\nonumber \\ & &
\left.
+
\left[
e Q_{\psi_{\mbox{\tiny down}}} (\partial_\nu A_{\mu})
+
\frac{g}{2 \cos \theta_W}
(T_{3,\psi_{\mbox{\tiny down},L}}-
2 Q_{\psi_{\mbox{\tiny down}}} \sin^2 \theta_W)
(\partial_\nu Z_{\mu})
\right.
\right.
\nonumber \\ & &
\left.
\left.
\quad -i \, \frac{g^2}{2} W_{\mu}^+ W_{\nu}^-
\right.
\Big]
\left( 1 + \frac{h}{v}\right)
\right\}
\, ,
\label{eq:NdV}
\end{eqnarray}
\begin{eqnarray}
N_{dd}^{A,\theta (ij)}
&=&
\frac{g}{4 \cos \theta_W}
\,
T_{3,\psi_{\mbox{\tiny down},L}}
\,
 \theta^{\mu \nu}
M_{\mbox{\tiny down}}^{(ij)}
\left( 1 + \frac{h}{v}\right)
Z_{\mu}
\nonumber \\ & & \times
\left[
\left(
\stackrel{\leftarrow}{\partial}_{\nu}
- \stackrel{\rightarrow}{\partial}_{\nu}
\right)
+2 i
e
Q_{\psi_{\mbox{\tiny down}}}
A_{\nu}
\right]\, , \qquad
\label{eq:NdA}
\end{eqnarray}
and
\begin{equation}
\left.
\begin{array}{l}
 N_{uu}^{V,\theta (ij)}\\[0.2cm]
N_{uu}^{A,\theta (ij)}
\end{array}
\right\}
 =
\left\{
\begin{array}{l}
N_{dd}^{V,\theta (ij)}\\[0.2cm]
N_{dd}^{A,\theta (ij)}
\end{array}
\right.
(
W^+  \leftrightarrow  W^-,
{\mbox{down}}
\rightarrow
\mbox{up}
 ).
\label{eq:NuVA}
\end{equation}
The charged currents are given by
\begin{eqnarray}
C_{ud}^{V(ij)}
&=&
C_{ud}^{V,\theta (ij)}
+
{\cal O}(\theta^2)
\, ,
\nonumber \\[0.1cm]
C_{ud}^{A(ij)}
&=&
C_{ud}^{A,\theta (ij)}
+
{\cal O}(\theta^2)
\, ,
\end{eqnarray}
where
\begin{eqnarray}
\lefteqn{C_{ud}^{V,\theta (ij)}}
\nonumber \\
&=&
- \frac{g}{4 \sqrt{2}}
\theta^{\mu \nu}
\left( 1 + \frac{h}{v}\right)
\left\{
\left[
\left(
(V_{f} M_{\mbox{\tiny down}})^{(ij)}+
(M_{\mbox{\tiny up}} V_{f})^{(ij)}
\right)
(\partial_{\nu} W^+_{\mu})
\right.
\right.
\nonumber \\[0.1cm] & &
\left.
\left.
+
\left(
(V_{f} M_{\mbox{\tiny down}})^{(ij)}
\stackrel{\rightarrow}{\partial}_{\nu}
+
(M_{\mbox{\tiny up}} V_{f})^{(ij)}
\stackrel{\leftarrow}{\partial}_{\nu}
\right)
W^+_{\mu}
\right]
\right.
\nonumber \\[0.1cm] & &
\left.
+ i e
\left(
(V_{f} M_{\mbox{\tiny down}})^{(ij)}
Q_{\psi_{\mbox{\tiny up}}}
-
(M_{\mbox{\tiny up}} V_{f})^{(ij)}
Q_{\psi_{\mbox{\tiny down}}}
\right)
A_{\mu} W_{\nu}^+
\right.
\nonumber \\[0.1cm] & &
\left.
+ i \frac{g}{\cos \theta_W}
\left[
(V_{f} M_{\mbox{\tiny down}})^{(ij)}
\left(
2 T_{3,\psi_{\mbox{\tiny up},L}}
-
Q_{\psi_{\mbox{\tiny up}}}
\sin^2 \theta_W
\right)
\right.
\right.
\nonumber \\[0.1cm] & &
\left.
\left.
- (M_{\mbox{\tiny up}} V_{f})^{(ij)}
\left(
2 T_{3,\psi_{\mbox{\tiny down},L}}
-
Q_{\psi_{\mbox{\tiny down}}}
\sin^2 \theta_W
\right)
\right]
Z_{\mu} W_{\nu}^+
\right\}
\, ,
\end{eqnarray}
and
\begin{equation}
C_{ud}^{A,\theta (ij)} =
C_{ud}^{V,\theta (ij)}(M_{\mbox{\tiny up}} \rightarrow - M_{\mbox{\tiny up}})
\, ,
\end{equation}
while
\begin{eqnarray}
C_{du}^{V (ij)} & = &
\left( C_{ud}^{V (ij)}
(\stackrel{\rightarrow}{\partial}
\leftrightarrow
\stackrel{\leftarrow}{\partial}
)\right) ^{\dagger}
\, ,
\nonumber \\
C_{du}^{A (ij)} & = & -
\left( C_{ud}^{A (ij)}
(\stackrel{\rightarrow}{\partial}
\leftrightarrow
\stackrel{\leftarrow}{\partial}
)\right) ^{\dagger}\, .
\label{eq:ChVA}
\end{eqnarray}
Note that $\stackrel{\rightarrow}{\partial}$
and $\stackrel{\leftarrow}{\partial}$
are defined in \req{eq:partial}.

At the end, observe that  the simplified
introduction of the fermion mass and the use of the relation
\begin{eqnarray}
S_{\mbox{\tiny $\psi$,m}} & = & \int d^4x \, \overline{\widehat \psi} \,
 *  (i \widehat{\fmslash D} -m)\, \widehat \psi
 \nonumber\\
& = & \int d^4x
\left[ \overline{\psi} \, (i \fmslash D -m)\, \psi
-\frac{1}{4}\,  \overline{\psi}\, \Re_{\psi}(F_{\mu \nu}) \,
 (i \theta^{\mu \nu \rho} \,  D_{\rho} -m \, \theta^{\mu \nu})
  \psi \nonumber \right. \\ & & \left.
  + {\cal O}(\theta^2) \right]
\, .
\label{eq:hatPsiDPsiM}
\end{eqnarray}
is valid only in the case of pure QED and pure QCD.

\section{Feynman Rules}
\label{app:ewFR}

On the basis of the results presented in Sections~\ref{sec:ew} and
\ref{sec:higgs}, it is now straightforward to derive the Feynman
rules needed for phenomenological applications of the NCSM, i.e.
for the calculation of physical processes. In this section, we list
a number of selected Feynman rules for the NCSM pure electroweak
interactions up to order $\theta$.
We omit interactions with the Higgs particle, boson interactions with
four and more gauge fields,
and fermion interactions with more than two gauge bosons.

The following notation for vertices has been adopted:
{\it all gauge boson momenta are taken to be incoming;
following the flow of the fermion line,
the momenta of the incoming and outgoing fermions
are given by $p_{\mbox{\tiny {\rm in}}}$ and
$p_{\mbox{\tiny {\rm out}}}$, respectively}.
In the following we denote fermions by $f$,
and the generation indices by $i$ and $j$.
Furthermore,
$f_u^{(i)} \in \{\nu^{(i)},u^{(i)}\}$
and
$f_d^{(i)} \in \{e^{(i)},d^{(i)}\}$.

For the Feynman rules we use the following definitions:
\begin{eqnarray}
c_{V,f} & = &
T_{3,f_{L}}\, -\,  2\,  Q_{f}\,  \sin^2 \theta_W\,,
\nonumber\\
c_{A,f} & = & T_{3,f_L}
\,.
\label{cva}
\end{eqnarray}
The charge $Q$ and the weak isospin $T_3$
can be read from Table \ref{tab:SMfields}.
The notation
$V_{f}$ is introduced in \req{eq:fCKM},
while $\theta^{\mu \nu \rho}$ is defined in (\ref{eq:theta3}).
We also make use of
$(\theta k)^\mu\equiv\theta^{\mu\nu}k_\nu
=- k_\nu\theta^{\nu\mu}\equiv -(k\theta)^\mu$
and $(k \theta p)\equiv k_\mu \theta^{\mu\nu} p_{\nu}$.


\subsection{Minimal NCSM}

In this subsection we present selected Feynman rules for the mNCSM
containing SM contributions and $\theta$ corrections. The $\theta$
corrections to vertices containing fermions are obtained using Eq.
(\ref{eq:Sew}) and the Yukawa part  of the action
\req{eq:SYukawa-fiz} has to be taken into account as well, because
it generates additional mass dependent terms which modify some
interaction vertices. In comparison with the SM, this is a novel
feature. Similarly, the gauge boson couplings present in
\req{eq:SmNCSMa} receive additional $\theta$ dependent corrections
from the Higgs part of the action \req{eq:H} and even new three- and
four-gauge boson couplings appear, see (\ref{eq:higgsWWA}).

First, we list three-vertices that appear in the SM as well.

\noindent
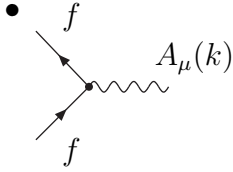
\begin{itemize}
\item \hspace{-1cm}\\
\begin{picture}(55,45) (30,-30)
\SetWidth{0.5}
\ArrowLine(30,-30)(50,-10)
\ArrowLine(50,-10)(30,11)
\Photon(80,-10)(50,-10){2}{4}
\Vertex(50,-10){1.5}
\Text(40,11)[lb]{$f$}
\Text(40,-40)[lb]{$f$}
\Text(75,-5)[lb]{$A_{\mu}(k)$}
\end{picture}
\\
\begin{eqnarray}
\lefteqn{i \, e \, Q_{f}\,
  \left[
 \gamma_{\mu}
 - \frac{i}{2} \, k^{\nu}
   \left( \theta_{\mu \nu \rho}
 \, p_{\mbox{\tiny in}}^{\rho}-
\theta_{\mu \nu}\, m_f\,\right)
  \right]}
\nonumber \\[0.2cm]
& = &
i \, e \, Q_{f}\,
\gamma_{\mu}
\nonumber \\
& &
+ \frac{1}{2} \, e \, Q_{f}\,
\left[
(p_{\mbox{\tiny out}} \theta p_{\mbox{\tiny in}}) \gamma_\mu
-
(p_{\mbox{\tiny out}}\theta)_\mu(\slash \!\!\! p_{\mbox{\tiny in}}-m_f)
-
(\slash \! \! \! p_{\mbox{\tiny out}}-m_f)(\theta p_{\mbox{\tiny in}})_\mu
\right]\,,
\nonumber \\
\label{eq:ffgamma}
\end{eqnarray}

\item\hspace{1cm}\\
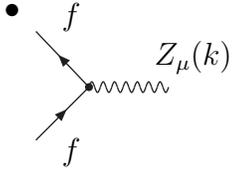
\begin{picture}(55,45) (30,-30)
\SetWidth{0.5}
\ArrowLine(30,-30)(50,-10)
\ArrowLine(50,-10)(30,11)
\Photon(80,-10)(50,-10){2}{7}
\Vertex(50,-10){1.5}
\Text(40,11)[lb]{$f$}
\Text(40,-40)[lb]{$f$}
\Text(75,-5)[lb]{$Z_{\mu}(k)$}
\end{picture}

\begin{eqnarray}
\hspace{1.0cm}
&&\frac{i \, e}{\sin2\theta_W} \,
   \Big\{\Big(
\gamma_{\mu}
 - \,
 \, \frac{i}{2}
 \, k^{\nu}
   \theta_{\mu \nu \rho}
 \, p_{\mbox{\tiny in}}^{\rho}\Big)
 \left( c_{V,f} - c_{A,f} \, \gamma_5 \right)
 \nonumber \\
 \hspace{-2cm}
 &&
\quad -
\, \frac{i}{2}\theta_{\mu \nu}\, m_f
  \Big[
     p_{\mbox{\tiny in}}^{\nu}\left( c_{V,f} - c_{A,f} \, \gamma_5 \right)
   -
     p_{\mbox{\tiny out}}^{\nu}\left( c_{V,f} + c_{A,f} \, \gamma_5 \right)
  \Big]
  \Big\} \,
  \,,
 \nonumber \\ & &
 \label{eq:ffZ}
\end{eqnarray}
\\

\item\hspace{1cm}\\
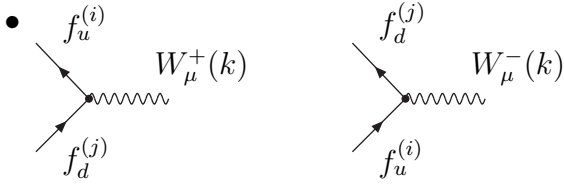
\begin{picture}(55,45) (30,-30)
\SetWidth{0.5}
\ArrowLine(30,-30)(50,-10)
\ArrowLine(50,-10)(30,11)
\Photon(80,-10)(50,-10){2}{7}
\Vertex(50,-10){1.5}
\Text(40,11)[lb]{$f_{u}^{(i)}$}
\Text(40,-40)[lb]{$f_{d}^{(j)}$}
\Text(75,-5)[lb]{$W^+_{\mu}(k)$}
\end{picture}
\hspace{2cm}
\begin{picture}(55,45) (30,-30)
\SetWidth{0.5}
\ArrowLine(30,-30)(50,-10)
\ArrowLine(50,-10)(30,11)
\Photon(80,-10)(50,-10){2}{7}
\Vertex(50,-10){1.5}
\Text(40,11)[lb]{$f_{d}^{(j)}$}
\Text(40,-40)[lb]{$f_{u}^{(i)}$}
\Text(75,-5)[lb]{$W^-_{\mu}(k)$}
\end{picture}
\vspace{1.0cm}
\begin{eqnarray}
\hspace{-.5cm}
&&
\hspace{-.5cm}
\frac{i \, e}{2 \sqrt{2} \sin\theta_W} \,
{V^{(ij)}_f \choose V^{*(ij)}_{f}}
  \left\{ \left[
\gamma_{\mu}
 -  \,
 \, \frac{i}{2} \, \theta_{\mu \nu \rho}
 \, k^{\nu}
 \, p_{\mbox{\tiny in}}^{\rho}
   \right] \,
 \left( 1 - \gamma_5 \right)
\right.
\nonumber
\\
\hspace{-.5cm}
&&
\hspace{-.5cm}
\left.
\quad -\frac{i}{2} \, \theta_{\mu \nu}
\left[
{m_{f_{u}^{(i)}} \choose m_{f_{d}^{(j)}}}
\, p_{\mbox{\tiny in}}^{\nu}(1-\gamma_5)
-
{m_{f_{d}^{(j)}} \choose m_{f_{u}^{(i)}}}
\, p_{\mbox{\tiny out}}^{\nu}(1+\gamma_5)
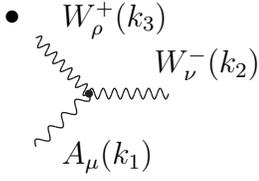
\right]
 \right\}
\,,
\label{eq:ffW}
\end{eqnarray}
\\

\item\hspace{1cm}\\
\begin{picture}(55,45) (30,-30)
\SetWidth{0.5}
\Photon(30,-30)(50,-10){2}{4}
\Photon(50,-10)(30,11){2}{7}
\Photon(80,-10)(50,-10){2}{7}
\Vertex(50,-10){1.5}
\Text(40,11)[lb]{$W^+_{\rho}(k_3)$}
\Text(40,-40)[lb]{$A_{\mu}(k_1)$}
\Text(75,-5)[lb]{$W^-_{\nu}(k_2)$}
\end{picture}

\begin{eqnarray}
\hspace{-.5cm}
&&
\hspace{-.5cm}
i e\;\Big\{g^{\mu\nu}(k_1-k_2)^\rho +
 g^{\nu\rho}(k_2-k_3)^\mu +
 g^{\rho\mu}(k_3-k_1)^\nu
\nonumber
\\
\hspace{-.5cm}
&&
\hspace{-.5cm}
\quad +\frac{i}{2} M_W^2 \Big[
\theta^{\mu \nu} k_1^\rho +
\theta^{\mu \rho} k_1^\nu
 + g^{\mu\nu}(\theta k_1)^\rho
-g^{\nu\rho}(\theta k_1)^\mu + g^{\rho\mu}(\theta k_1)^\nu \Big]\Big\}
\nonumber \,,
\\ & &
\label{2Wgamma}
\end{eqnarray}
\\

\item\hspace{1cm}\\
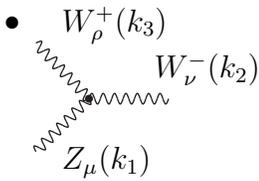
\begin{picture}(55,45) (30,-30)
\SetWidth{0.5}
\Photon(30,-30)(50,-10){2}{7}
\Photon(50,-10)(30,11){2}{7}
\Photon(80,-10)(50,-10){2}{7}
\Vertex(50,-10){1.5}
\Text(40,11)[lb]{$W^+_{\rho}(k_3)$}
\Text(40,-40)[lb]{$Z_{\mu}(k_1)$}
\Text(75,-5)[lb]{$W^-_{\nu}(k_2)$}
\end{picture}

\begin{eqnarray}
\hspace{-.5cm}
&&
\hspace{-1.0cm}
i e\;{\cot \theta_W}\Big\{g^{\mu\nu}(k_1-k_2)^\rho +
 g^{\nu\rho}(k_2-k_3)^\mu +
 g^{\rho\mu}(k_3-k_1)^\nu
\nonumber
\\
\hspace{-.5cm}
&&
\hspace{-.5cm}
+\frac{i}{2} M_W^2
\Big[ \theta^{\mu \nu} k_1^\rho +
\theta^{\mu \rho} k_1^\nu
 + g^{\mu\nu}(\theta k_1)^\rho
-g^{\nu\rho}(\theta k_1)^\mu + g^{\rho\mu}(\theta k_1)^\nu
\Big]
\nonumber\\
\hspace{-.5cm}
&&
\hspace{-.5cm}
-
\frac{i}{4}
{M_Z^2}
\Big[
\theta^{\mu \nu} (k_1-k_2)^\rho
+
\theta^{\nu \rho} (k_2-k_3)^\mu
+
\theta^{\rho \mu} (k_3-k_1)^\nu
\nonumber \\
\hspace{-.5cm}
&&
\hspace{-.5cm}
 - 2 g^{\mu\nu}(\theta k_3)^\rho
 - 2 g^{\nu \rho}(\theta k_1)^\mu
 - 2 g^{\rho \mu}(\theta k_2)^\nu
\Big] \Big\} \,.
\label{WWZ}
\end{eqnarray}

\end{itemize}

Here we give the new three-gauge-boson coupling
which follows from the Higgs action \req{eq:H},
i.e., Eq. \req{eq:higgsWWA}
\begin{itemize}
\item\hspace{1cm}\\
\begin{picture}(55,45) (30,-30)
\SetWidth{0.5}
\Photon(30,-30)(50,-10){2}{7}
\Photon(50,-10)(30,11){2}{7}
\Photon(80,-10)(50,-10){2}{7}
\Vertex(50,-10){1.5}
\Text(40,11)[lb]{$Z_{\rho}(k_3)$}
\Text(40,-40)[lb]{$Z_{\mu}(k_1)$}
\Text(75,-5)[lb]{$Z_{\nu}(k_2)$}
\end{picture}
\\

\vspace{-2cm}
\begin{eqnarray}
\displaystyle
\hspace{2.5cm}
&&
\hspace{-.5cm}
\frac{e\,M_Z^2}{2 \sin2\theta_W}
\Big[
\theta^{\mu \nu} (k_1-k_2)^\rho
+
\theta^{\nu \rho} (k_2-k_3)^\mu
+
\theta^{\rho \mu} (k_3-k_1)^\nu
\nonumber \\
\hspace{-.5cm}
&&
\hspace{-.5cm}
 - 2 g^{\mu\nu}(\theta k_3)^\rho
 - 2 g^{\nu \rho}(\theta k_1)^\mu
 - 2 g^{\rho \mu}(\theta k_2)^\nu
\Big] \,.
\label{eq:ZZZm}
\end{eqnarray}
\\
\end{itemize}
Additionally,
from the Higgs action \req{eq:H}
one can derive
the $\theta$ corrections to the electroweak four-gauge-boson vertices
already present in SM (see \req{eq:SmNCSMa}),
as well as, new four-gauge-boson vertices.

Equation (\ref{eq:Sew}) also describes the interaction vertices
involving fermions and two or three gauge bosons. These do not
appear in the SM. In the following we provide all contributions to
such vertices with four legs and corresponding mass-dependent
contributions from \req{eq:SYukawa-fiz}.
\\
\begin{itemize}

\item \hspace{1cm}\\
\begin{picture}(55,45) (30,-30)
\SetWidth{0.5}
\ArrowLine(30,-30)(50,-10)
\ArrowLine(50,-10)(30,11)
\Photon(70,-30)(50,-10){2}{4}
\Photon(70,11)(50,-10){2}{4}
\Vertex(50,-10){1.5}
\Text(40,11)[lb]{$f$}
\Text(40,-40)[lb]{$f$}
\Text(77,-40)[lb]{$A_{\nu}(k_2)$}
\Text(77,8)[lb]{$A_{\mu}(k_1)$}
\end{picture}
\vspace{-1.5cm}
\begin{eqnarray}
\frac{-\,e^2\, Q^2_f}{2} \,
    \theta_{\mu\nu\rho} \, (k_1^{\rho}-k_2^{\rho}) \,,
\label{eq:ffgammagamma}
\end{eqnarray}\\

\item \hspace{1cm}
\\

\begin{picture}(55,45) (30,-30)
\SetWidth{0.5}
\ArrowLine(30,-30)(50,-10)
\ArrowLine(50,-10)(30,11)
\Photon(70,-30)(50,-10){2}{7}
\Photon(70,11)(50,-10){2}{4}
\Vertex(50,-10){1.5}
\Text(40,11)[lb]{$f$}
\Text(40,-40)[lb]{$f$}
\Text(77,-40)[lb]{$Z_{\nu}(k_2)$}
\Text(77,8)[lb]{$A_{\mu}(k_1)$}
\end{picture}
\\

\vspace{-3.5cm}
\begin{eqnarray}
&&
\hspace{2.5cm}
\frac{-\,e^2 \, Q_f }{2\sin2\theta}
\label{eq:ffgammaZ}\\
&&
\hspace{2.5cm}
\times
\Big[\theta_{\mu\nu\rho}\,(k_1^{\rho}-k_2^{\rho})(c_{V,f}-c_{A,f}\gamma_5)
-2\theta_{\mu\nu}\,m_f\,c_{A,f}\gamma_5\Big] \,,
\nonumber
\end{eqnarray}\\

\item \hspace{1cm}\\
\begin{picture}(55,45) (30,-30)
\SetWidth{0.5}
\ArrowLine(30,-30)(50,-10)
\ArrowLine(50,-10)(30,11)
\Photon(70,-30)(50,-10){2}{7}
\Photon(70,11)(50,-10){2}{7}
\Vertex(50,-10){1.5}
\Text(40,11)[lb]{$f$}
\Text(40,-40)[lb]{$f$}
\Text(77,-40)[lb]{$Z_{\nu}(k_2)$}
\Text(77,8)[lb]{$Z_{\mu}(k_1)$}
\end{picture}
\vspace{-1.5cm}
\begin{eqnarray}
\hspace{2.5cm}
\frac{-\,e^2}{2 \sin^2 2\theta} \,
    \theta_{\mu\nu\rho} \,(k_1^{\rho}-k_2^{\rho})\,
    (c_{V,f}-c_{A,f}\gamma_5)^2 \,,
\label{eq:ffZZ}
\end{eqnarray}\\

\item \hspace{1cm}\\
\begin{picture}(55,45) (30,-30)
\SetWidth{0.5}
\ArrowLine(30,-30)(50,-10)
\ArrowLine(50,-10)(30,11)
\Photon(70,-30)(50,-10){2}{7}
\Photon(70,11)(50,-10){2}{7}
\Vertex(50,-10){1.5}
\Text(40,11)[lb]{$f$}
\Text(40,-40)[lb]{$f$}
\Text(77,-40)[lb]{$W^-_{\nu}(k_2)$}
\Text(77,8)[lb]{$W^+_{\mu}(k_1)$}
\end{picture}
\vspace{-1.5cm}
\begin{eqnarray}
\hspace{3cm}
\frac{-\,e^2}{8\sin^2\theta_W} \,
\Big[ \theta_{\mu\nu\rho} \,(p_{{\mbox{\tiny in}}}^{\rho}+k_1^{\rho})
    (1-\gamma_5) + 2\theta_{\mu\nu} m_f\Big] \,,
\label{eq:ffWW}
\end{eqnarray}
\vspace{.5cm}

\item\hspace{1cm}\\
\hspace*{1cm}
\begin{picture}(55,45) (30,-30)
\SetWidth{0.5}
\ArrowLine(30,-30)(50,-10)
\ArrowLine(50,-10)(30,11)
\Photon(70,-30)(50,-10){2}{7}
\Photon(70,11)(50,-10){2}{4}
\Vertex(50,-10){1.5}
\Text(5,11)[lb]{$f_{u}^{(i)}$}
\Text(3,-40)[lb]{$f_{d}^{(j)}$}
\Text(77,-40)[lb]{$W^+_{\nu}(k_2)$}
\Text(77,8)[lb]{$A_{\mu}(k_1)$}
\end{picture}
\hspace*{3cm}
\begin{picture}(55,45) (30,-30)
\SetWidth{0.5}
\ArrowLine(30,-30)(50,-10)
\ArrowLine(50,-10)(30,11)
\Photon(70,-30)(50,-10){2}{7}
\Photon(70,11)(50,-10){2}{4}
\Vertex(50,-10){1.5}
\Text(3,11)[lb]{$f_{d}^{(j)}$}
\Text(5,-40)[lb]{$f_{u}^{(i)}$}
\Text(77,-40)[lb]{$W^-_{\nu}(k_2)$}
\Text(77,8)[lb]{$A_{\mu}(k_1)$}
\end{picture}

\begin{eqnarray}
&&
\hspace{-1cm}
\frac{- e^2}{4 \sqrt{2} \sin\theta_W}
\left\{ \theta_{\mu \nu\rho}
\left[
{
    Q_{f_{u}^{(i)}} \choose
    Q_{f_{d}^{(j)}}
}
(p_{{\mbox{\tiny in}}}^{\rho}+k_1^{\rho})
-
{
    Q_{f_{d}^{(j)}} \choose
    Q_{f_{u}^{(i)}}
}
(p_{{\mbox{\tiny in}}}^{\rho} + k_2^{\rho})
   \right] \,
 \left( 1 - \gamma_5 \right)
 \right.
\nonumber
\\
&&
\hspace{-1.cm}
\left.
+\; \theta_{\mu \nu}
\left[
{
m_{f^{(i)}_{u}}
Q_{f_{\mbox{\tiny d}}^{(j)}} \choose
m_{f^{(j)}_{d}} Q_{f_{\mbox{\tiny u}}^{(i)}}
}
(1-\gamma_5)
-
{
m_{f^{(j)}_{d}}
  Q_{f_{u}^{(i)}}\choose
m_{f^{(i)}_{u}}
  Q_{f_{d}^{(j)}}
}
(1+\gamma_5)
\right]
 \right\}
{
V^{(ij)}_f \choose
V^{*(ij)}_{f}
} \,,
\nonumber\\
\label{eq:qqgammaW}
\end{eqnarray}
%

\item \hspace{1cm}\\
\hspace*{1cm}
\begin{picture}(55,45) (30,-30)
\SetWidth{0.5}
\ArrowLine(30,-30)(50,-10)
\ArrowLine(50,-10)(30,11)
\Photon(70,-30)(50,-10){2}{7}
\Photon(70,11)(50,-10){2}{7}
\Vertex(50,-10){1.5}
\Text(5,11)[lb]{$f_{u}^{(i)}$}
\Text(3,-40)[lb]{$f_{d}^{(j)}$}
\Text(77,-40)[lb]{$W^+_{\nu}(k_2)$}
\Text(77,8)[lb]{$Z_{\mu}(k_1)$}
\end{picture}
\hspace*{3cm}
\begin{picture}(55,45) (30,-30)
\SetWidth{0.5}
\ArrowLine(30,-30)(50,-10)
\ArrowLine(50,-10)(30,11)
\Photon(70,-30)(50,-10){2}{7}
\Photon(70,11)(50,-10){2}{7}
\Vertex(50,-10){1.5}
\Text(3,11)[lb]{$f_{d}^{(j)}$}
\Text(5,-40)[lb]{$f_{u}^{(i)}$}
\Text(77,-40)[lb]{$W^-_{\nu}(k_2)$}
\Text(77,8)[lb]{$Z_{\mu}(k_1)$}
\end{picture}

\begin{eqnarray}
&&
\hspace{-1.7cm}
\frac{-\,e^2 }{4  \sqrt{2} \sin\theta_W \sin2\theta_W}
{
V^{(ij)}_f \choose
V^{*(ij)}_{f}
}
 \nonumber
\\[0.1cm]
&&
\hspace{-1.7cm}
 \left\{
    \theta_{\mu\nu\rho}
   \left[
   \left({
     c_{V,f^{(i)}_{u}}+c_{A,f^{(i)}_{u}} \atop
     c_{V,f^{(j)}_{d}}+c_{A,f^{(j)}_{d}}
   }\right)
     (p_{{\mbox{\tiny in}}}^{\rho}+k_1^{\rho})
     \right.
 \left.
 - \left({
     c_{V,f^{(j)}_{d}}+c_{A,f^{(j)}_{d}} \atop
     c_{V,f^{(i)}_{u}}+c_{A,f^{(i)}_{u}}
    }\right)
     (p_{{\mbox{\tiny in}}}^{\rho} + k_2^{\rho})
   \right] \,
 \left( 1 - \gamma_5 \right)
 \right.
\nonumber
\\
 &&
 \left.
 +\;\theta_{\mu\nu}\left[
 \left({
 m_{f^{(i)}_{u}}
  \left[c_{V,f^{(j)}_{d}}+3c_{A,f^{(j)}_{d}}\right]
 \atop
 m_{f^{(j)}_{d}}
  \left[c_{V,f^{(i)}_{u}}+3c_{A,f^{(i)}_{u}}\right]
 }\right)
 \left( 1 - \gamma_5 \right)
\right.
\right.
\nonumber \\[0.1cm]
& & \left.
\left.
\hspace{1.cm}
 -
 \left({
 m_{f^{(j)}_{d}}
\left[c_{V,f^{(i)}_{u}}+3c_{A,f^{(i)}_{u}}\right]
 \atop
 m_{f^{(i)}_{u}}
\left[c_{V,f^{(j)}_{d}}+3c_{A,f^{(j)}_{d}}\right]
 }\right)
 \left( 1 + \gamma_5 \right)
 \right]
 \right\}\,.
 \label{eq:qqZW}
\end{eqnarray}
\end{itemize}
Similarly, $ffWWZ$, $ffWW\gamma$ and $ff\gamma WZ$ can be extracted
from Eq. (\ref{eq:Sew}) as well. They have no mass-dependent
corrections.

\subsection{\label{sec:FRnon-minimal}Non-Minimal NCSM}

Here we give the selected Feynman rules for the non-minimal
NCSM introduced in Subsection \ref{sec:non-minimal}.
Observe that the fermion sector is not affected by the change of
the representation in the gauge part of the action.
Let us define
\begin{eqnarray}
& & \Theta_3((\mu,k_1),(\nu,k_2),(\rho,k_3)) =
-\theta^{\mu \nu}
[ k_1^\rho(k_2 k_3) - k_2^\rho (k_1 k_3) ]
\nonumber \\
& &
+
(\theta k_1)^\mu[g^{\nu \rho} (k_2 k_3) - k_2^\rho k_3^\nu]
-
(\theta k_1)^\nu[g^{\rho \mu} (k_2 k_3) - k_2^\rho k_3^\mu]
\nonumber \\
& &
-
(\theta k_1)^\rho[g^{\mu \nu} (k_2 k_3) - k_2^\mu k_3^\nu]
+
(k_1 \theta k_2)
[k_3^\mu g^{\nu \rho} - k_3^\nu g^{\rho \mu}]
\nonumber \\
& &
+
\mbox{cyclical permutations of $(\mu_i,k_i)$}
\, .
\label{eq:Gamma}
\end{eqnarray}
We use the simplified notation $\mu_1\equiv \mu$, $\mu_2\equiv \nu$
and $\mu_3\equiv \rho$.

First, we list the
Feynman rules for the modified $W^+W^-\gamma$, $W^+W^-Z$
and $ZZZ$ vertices already present in the mNCSM:
\noindent
\begin{itemize}
\item\hspace{1cm}\\
\begin{picture}(55,45) (30,-30)
\SetWidth{0.5}
\Photon(30,-30)(50,-10){2}{4}
\Photon(50,-10)(30,11){2}{7}
\Photon(80,-10)(50,-10){2}{7}
\Vertex(50,-10){1.5}
\Text(40,11)[lb]{$W^+_{\rho}(k_3)$}
\Text(40,-40)[lb]{$A_{\mu}(k_1)$}
\Text(75,-5)[lb]{$W^-_{\nu}(k_2)$}
\end{picture}
\\
\vspace{-2.cm}
\begin{eqnarray}
&&
\hspace{1.5cm}
{\rm Eq.}\; (\ref{2Wgamma}) +
\nonumber
\\
&&
\hspace{1.5cm}
2\,e\,\sin{2\theta_W}{\rm K}_{WW\gamma}  \,
\Theta_3((\mu,k_1),(\nu,k_2),(\rho,k_3))\,,
\label{nm2Wgamma}
\end{eqnarray}

\item\hspace{1cm}\\
\begin{picture}(55,45) (30,-30)
\SetWidth{0.5}
\Photon(30,-30)(50,-10){2}{7}
\Photon(50,-10)(30,11){2}{7}
\Photon(80,-10)(50,-10){2}{7}
\Vertex(50,-10){1.5}
\Text(40,11)[lb]{$W^+_{\rho}(k_3)$}
\Text(40,-40)[lb]{$Z_{\mu}(k_1)$}
\Text(75,-5)[lb]{$W^-_{\nu}(k_2)$}
\end{picture}
\\
\vspace{-2.cm}
\begin{eqnarray}
&&
\hspace{1.5cm}
{\rm Eq.}\; (\ref{WWZ}) +
\nonumber
\\
&&
\hspace{1.5cm}
2\,e\,\sin{2\theta_W}{\rm K}_{WWZ}  \,
\Theta_3((\mu,k_1),(\nu,k_2),(\rho,k_3))\,,
\label{nm2WZ}
\end{eqnarray}

\item\hspace{1cm}\\
\begin{picture}(55,45) (30,-30)
\SetWidth{0.5}
\Photon(30,-30)(50,-10){2}{7}
\Photon(50,-10)(30,11){2}{7}
\Photon(80,-10)(50,-10){2}{7}
\Vertex(50,-10){1.5}
\Text(40,11)[lb]{$Z_{\rho}(k_3)$}
\Text(40,-40)[lb]{$Z_{\mu}(k_1)$}
\Text(75,-5)[lb]{$Z_{\nu}(k_2)$}
\end{picture}
\\
\vspace{-2.cm}
\begin{eqnarray}
&&
\hspace{1.5cm}
{\rm Eq.}\; (\ref{eq:ZZZm}) +
\nonumber
\\
&&
\hspace{1.5cm}
2\,e\,\sin{2\theta_W}{\rm K}_{ZZZ}  \,
\Theta_3((\mu,k_1),(\nu,k_2),(\rho,k_3))\,.
\label{3Z}
\end{eqnarray}
\\
\end{itemize}
Additionally, we give the new gauge boson vertices
$\gamma\gamma\gamma$, $Z\gamma\gamma$
and $ZZ\gamma$:

\begin{itemize}
\item \hspace{2cm}\\
\begin{picture}(55,45) (30,-30)
\SetWidth{0.5}
\Photon(30,-30)(50,-10){2}{4}
\Photon(50,-10)(30,11){2}{4}
\Photon(80,-10)(50,-10){2}{4}
\Vertex(50,-10){1.5}
\Text(40,11)[lb]{$A_{\rho}(k_3)$}
\Text(40,-40)[lb]{$A_{\mu}(k_1)$}
\Text(75,-5)[lb]{$A_{\nu}(k_2)$}
\end{picture}
\\
\vspace{-2.cm}
\begin{equation}
\hspace{1.5cm}
2\,e\,\sin{2\theta_W}{\rm K}_{\gamma\gamma\gamma}  \,
\Theta_3((\mu,k_1),(\nu,k_2),(\rho,k_3))\,,
\label{3gamma}
\end{equation}
\\

\item\hspace{1cm}\\
\begin{picture}(55,45) (30,-30)
\SetWidth{0.5}
\Photon(30,-30)(50,-10){2}{4}
\Photon(50,-10)(30,11){2}{4}
\Photon(80,-10)(50,-10){2}{7}
\Vertex(50,-10){1.5}
\Text(40,11)[lb]{$A_{\rho}(k_3)$}
\Text(40,-40)[lb]{$A_{\mu}(k_1)$}
\Text(75,-5)[lb]{$Z_{\nu}(k_2)$}
\end{picture}
\\
\vspace{-2.cm}
\begin{equation}
\hspace{1.5cm}
-2\,e\,\sin{2\theta_W}{\rm K}_{Z\gamma\gamma}  \,
\Theta_3((\mu,k_1),(\nu,k_2),(\rho,k_3))\,,
\label{Z2gamma}
\end{equation}
\\

\item\hspace{1cm}\\
\begin{picture}(55,45) (30,-30)
\SetWidth{0.5}
\Photon(30,-30)(50,-10){2}{4}
\Photon(50,-10)(30,11){2}{7}
\Photon(80,-10)(50,-10){2}{7}
\Vertex(50,-10){1.5}
\Text(40,11)[lb]{$Z_{\rho}(k_3)$}
\Text(40,-40)[lb]{$A_{\mu}(k_1)$}
\Text(75,-5)[lb]{$Z_{\nu}(k_2)$}
\end{picture}
\\
\vspace{-2.cm}
\begin{equation}
\hspace{1.5cm}
-2\,e\,\sin{2\theta_W}{\rm K}_{ZZ\gamma}  \,
\Theta_3((\mu,k_1),(\nu,k_2),(\rho,k_3))\,.
\label{2Zgamma}
\end{equation}
\\

\end{itemize}
The functions $K$ are not independent
and they are defined in Eqs. (\ref{K123456},\ref{WWgammaZ}).

\section{\label{sec:con}Conclusions}

The main purpose of this article is to complete the Non-Commutative
Standard Model constructed in \cite{Calmet:2001na,Behr:2002wx}, and
thus to make it accessible to phenomenological considerations and
further research. 
The NCSM action are given in terms of 
physical fields and mass eigenstates. 
The freedom in the choice of traces in
kinetic terms for gauge fields produces two versions of the NCSM, 
namely the
mNCSM and the nmNCSM. However, such freedom does not affect the
matter sector of the action and the fermion-gauge boson interactions
remain the same in both versions of the NCSM. We have provided an
explicit expression for selected vertices of which some already
appear in the original SM, but in the NCSM they gain
$\theta$-dependent corrections, whereas others appear for the first
time in the non-commutative version of the SM. We have presented a
careful discussion of electroweak charged and neutral currents as
well as a derivation of the Higgs and Yukawa terms of the
NCSM action.

Among the novel features in comparison with previous works
\cite{Calmet:2001na,Behr:2002wx} are the appearance of additional
gauge boson interaction terms (\ref{LWW}) and (\ref{eq:higgsWWA}) in
the gauge (\ref{eq:genSgaugeSM}) and in the Higgs (\ref{eq:H}) parts
of the action, and the appearance of mass-dependent corrections to
the boson-boson and fermion-boson couplings steaming from the Higgs
and Yukawa parts of the action, respectively.
In Eqs. (\ref{eq:ffgamma}-\ref{eq:ffW}) the mass-dependent terms
stem from the Yukawa interactions (\ref{eq:Sy}-\ref{eq:ChVA}), while
in Eqs. (\ref{2Wgamma},\ref{WWZ}) the mass corrections arrise from
the $\theta$-expanded Higgs action (\ref{eq:higgsWWA}).
To first order in $\theta$, equation (\ref{eq:SYukawa-fiz}) contains
coupling of fermions to gauge bosons that depend on the mass
of the fermion involved. Also the appearance of new terms
(\ref{eq:higgsWWA}) would certainly produce important contributions
in a number of physical processes. All the above features are introduced
by the Seiberg-Witten maps.

CP violation induced by space-time non-commutativity has potential
to be a particular sensitive probe of non-commutativity
\cite{Hinchliffe:2001im}. The analysis of C,P,T properties of the
NCSM and the NCGUTs \cite{Aschieri:2002mc} shows that $\theta$
transforms under C, P, T in such a way that it preserves these
discrete symmetries in the action. However, considering $\theta$ as
a fixed background (or spectator) field, there will be spontaneous
breaking of CP (relative to the background), just as one has
spontaneous breaking of Lorentz symmetries in non-commutative
theories. Consequently, non-commutative effects can also mix with
the CKM-matrix CP-violating parameter $\delta$ in the spirit of Ref.
\cite{Hinchliffe:2001im}. Since the fermion sectors of the mNCSM and
the nmNCS are equal, the above conclusion is valid for both models.
It should be noted that in the present work, 
the unitary CKM mixing matrix has been
considered with matrix elements not as functions of space-time but
as constants. Furthermore, the $\theta$-expansions
of the SW map and the star product have been worked out only up to
first order.

In conclusion, the thorough analysis of the electroweak sector
considered in this paper facilitates further research on reliable
bounds on non-commutativity from hadronic and leptonic physics.

\subsection*{Acknowledgment}

We want to thank Xavier Calmet, Goran Duplan\v{c}i\'{c}, Lutz
M\"oller and Julius Wess for many fruitful discussions. M.W. also
wants to acknowledge the support from ``Fonds zur F\"orderung der
wissenschaftlichen Forschung" (Austrian Science Fund), projects
P15463-N08 and P16779-N02 and from Ludwig-Maximillians-Universit\"at,
M\"unchen, Sektion Physik. This work was partially supported by the
Ministry of Science, Education and Sport of the Republic of Croatia.


\begin{thebibliography}{10}

\bibitem{Kontsevich:1997vb}
M.~Kontsevich,
Lett.\ Math.\ Phys.\  {\bf 66} (2003) 157
[q-alg/9709040].

\bibitem{Seiberg:1999vs}
N. Seiberg and E. Witten,
JHEP {\bf 09} (1999) 032 [hep-th/9908142].

\bibitem{Madore:2000en}
J. Madore, S. Schraml, P.~Schupp and J.~Wess,
Eur. Phys. J. C{\bf 16} (2000) 161 [hep-th/0001203].

\bibitem{Jurco:2000ja}
B. Jur\v{c}o, S. Schraml, P.~Schupp and J.~Wess,
Eur. Phys. J. C{\bf 17} (2000) 521 [hep-th/0006246].

\bibitem{Jurco:2001rq}
B. Jur\v{c}o, L. M\"oller, S.~Schraml, P.~Schupp and J.~Wess,
Eur. Phys. J. C{\bf 21} (2001) 383 [hep-th/0104153].

\bibitem{Calmet:2001na}
X.~Calmet, B.~Jur\v{c}o, P.~Schupp, J.~Wess and M.~Wohlgenannt,
Eur.~Phys.~J. C{\bf 23} (2002) 363
[hep-ph/0111115].

\bibitem{Chaichian:2004yh}
M.~Chaichian, P.~Presnajder and A.~Tureanu,
hep-th/0409096; to appear in Phys. Rev. Lett.

\bibitem{Koch:2004ud}
F.~Koch and E.~Tsouchnika,
hep-th/0409012.


\bibitem{Behr:2002wx}
W.~Behr, N.G. Deshpande, G. ~Duplan\v{c}i\'{c}, P.~Schupp, J.~Trampeti\'{c} and
J.~Wess,
Eur. Phys. J. C{\bf 29} (2003) 441 [hep-ph/0202121].

\bibitem{Duplancic:2003hg}
G.~Duplan\v{c}i\'{c}, P.~Schupp and J.~Trampeti\'{c},
Eur.~Phys. J. C{\bf 32} (2003) 141 [hep-ph/0309138].

\bibitem{Schupp:2002up}
P.~Schupp, J.~Trampeti\'{c}, J.~Wess and G.~Raffelt,
Eur.\ Phys.\ J.\ C {\bf 36} (2004) 405
[hep-ph/0212292].

\bibitem{Minkowski:2003jg}
P.~Minkowski, P.~Schupp and J.~Trampeti\'{c},
Eur.\ Phys.\ J.\ C {\bf 37} (2004) 123
[hep-th/0302175].

\bibitem{Trampetic:2002eb}
J. Trampeti\'c,
Acta Phys. Polon. B{\bf 33}
(2002) 4317 [hep-ph/0212309].

\bibitem{Schupp:2004dz}
P. Schupp and J. Trampeti\'c,
hep-ph/0405163.

\bibitem{Ohl:2004tn}
T.~Ohl and J.~Reuter,
Phys. Rev. D{\bf 70} (2004) 076007 [hep-ph/0406098].

\bibitem{Armoni:2000xr} A.~Armoni,
Nucl. Phys. B{\bf 593} (2001) 229
[hep-th/0005208].

\bibitem{Bichl:2001cq}
A.~Bichl, J. Grimstrup, H. Grosse, L. Popp, M. Schweda and R. Wulkenhaar,
JHEP {\bf 06} (2001) 013 [hep-th/0104097].

\bibitem{Grimstrup:2002xs}
J. Grimstrup, B. Kloibock, L. Popp, V. Putz, M. Schweda and M. Wickenhauser,
hep-th/0210288.

\bibitem{Brandt:2003fx}
F. Brandt, C.P. Mart\'{i}n and F. Ruiz Ruiz,
JHEP {\bf 07} (2003) 068
[hep-th/0307292].

\bibitem{Gomis:2000zz}
J. Gomis and Th. Mehen,
Nucl. Phys. B{\bf 591} (2000) 265 [hep-th/0005129].

\bibitem{Calmet:2003jv}
X. Calmet and M. Wohlgenannt,
Phys. Rev. D{\bf 68} (2003) 025016 [hep-ph/0305027].

\bibitem{Dimitrijevic:2003wv}
M.~Dimitrijevi\'{c}, L.~Jonke, L.~M\"oller, E.~Tschouchnika, J.~Wess and
M.~Wohlgenannt,
Eur. Phys. J. C{\bf 31} (2003) 129 [hep-th/0307149].

\bibitem{Dimitrijevic:2003pn}
M.~Dimitrijevi\'{c}, F.~Meyer, L.~M\"oller and J.~Wess,
Eur. Phys. J. C{\bf 36} (2004) 117 [hep-th/0310116].


\bibitem{Behr:2003qc}
W.~Behr and A.~Sykora,
Nucl. Phys. B{\bf 698} (2004) 473 [hep-th/0309145].

\bibitem{Grimstrup:2003rd}
J.M. Grimstrup, T. Jonsson and L. Thorlacius,
JHEP {\bf 12} (2003) 001 [hep-th/0310179].

\bibitem{Chaichian:2004za}
M.~Chaichian, P.~P.~Kulish, K.~Nishijima and A.~Tureanu,
Phys. Lett. B{\bf 604} (2004) 98 [hep-th/0408069].

\bibitem{Chaichian:2001aa}
M.~Chaichian, P.~Presnajder, M.M.~Sheikh-Jabbari and A.~Tureanu,
Eur.~Phys. J. C{\bf 29} (2003) 413 [hep-th/0107055].

\bibitem{Chaichian:2004yw}
M.~Chaichian, A.~Kobakhidze and A.~Tureanu,
hep-th/0408065.






\bibitem{Melic:2005am}
B.~Melic, K.~Passek-Kumericki, J.~Trampetic, P.~Schupp and M.~Wohlgenannt,
hep-ph/0503064.

\bibitem{Denk:2004um}
S. Denk, V.~Putz and M.~Wohlgenannt,
hep-th/0402229.

\bibitem{Aschieri:2002mc}
P. Aschieri, B.~Jur\v{c}o, P.~Schupp and J.~Wess,
Nucl. Phys. B{\bf 651} (2003) 45 [hep-th/0205214].

\bibitem{Weinberg:1995mt}
S.~Weinberg, ``The Quantum theory of fields. Vol. 1: Foundations,''
Section 7.2, Cambridge University Press 2000.


\bibitem{Hinchliffe:2001im}
I.~Hinchliffe and N.~Kersting,
Phys.\ Rev.\ D {\bf 64} (2001) 116007 [hep-ph/0104137].





\end{thebibliography}
\end{document}